\documentclass[journal,twoside,web]{ieeecolor}
\usepackage{tmi}
\usepackage{afterpage}
\usepackage{makecell}
\usepackage{amsmath}
\usepackage{amssymb}
\usepackage{amsfonts}
\usepackage{algorithmic}
\usepackage{graphicx}
\usepackage{textcomp}
\usepackage{siunitx}
\usepackage[table,xcdraw]{xcolor}
\usepackage[table]{xcolor}
\usepackage{multirow}
\usepackage{float}
\usepackage{tabularx} 
\usepackage{hyperref}
\usepackage{adjustbox}

\hypersetup{
hidelinks,
colorlinks=true,
linkcolor=red,
citecolor=cyan
}
\usepackage[noabbrev]{cleveref}
\usepackage{amsmath}

\crefname{table}{}{}

\usepackage{subfigure}

\renewcommand{\thetable}{\Roman{table}}
\def\BibT¬eX{{\rm B\kern-.05em{\sc i\kern-.025em b}\kern-.08em
    T\kern-.1667em\lower.7ex\hbox{E}\kern-.125emX}}

\begin{document} 
\title{Rethinking Brain Tumor Segmentation from the Frequency Domain Perspective}
\author{Minye Shao,
        Zeyu Wang$^{\dag, *}$,
       Haoran Duan,
       Yawen Huang$^{*}$,
       Bing Zhai,
       Shizheng Wang$^{*}$,\\
       Yang Long$^{*}$,
       Yefeng Zheng \textit{Fellow IEEE}
\thanks{Minye Shao and Yang Long are with the Department of Computer Science, Durham University. (E-mail: \{minye.shao, yang.long\}@ieee.org).}
\thanks{Zeyu Wang is with College of Computer Science and Engineering, Dalian Minzu University, China. (E-mail: 20231578@dlnu.edu.cn).}
\thanks{Haoran Duan is with the Department of Automation, Tsinghua University, China. (E-mail: haoran.duan@ieee.org).}
\thanks{Yawen Huang and Yefeng Zheng are with Jarvis Research Center, Tencent YouTu Lab, Shenzhen, China. (E-mail: \{yawenhuang, yefengzheng\}@tencent.com).} 
\thanks{Bing Zhai is with the School of Computing, Northumbria University, UK. (E-mail: bing.zhai@northumbria.ac.uk).}
\thanks{Shizheng Wang is with SunwayAI Research Lab, Fuyang Normal University, Fuyang, and also with Chinese Academy of Sciences R\&D Center for Internet of Things, China. (E-mail: shizheng.wang@foxmail.com).}
\thanks{$^{\dag}$ Co-first author.}
\thanks{$^{*}$ Corresponding authors.}
}
\maketitle
\begin{abstract}
Precise segmentation of brain tumors, particularly contrast-enhancing regions visible in post-contrast MRI (areas highlighted by contrast agent injection), is crucial for accurate clinical diagnosis and treatment planning but remains challenging. However, current methods exhibit notable performance degradation in segmenting these enhancing brain tumor areas, largely due to insufficient consideration of MRI-specific tumor features such as complex textures and directional variations. To address this, we propose the Harmonized Frequency Fusion Network (HFF-Net), which rethinks brain tumor segmentation from a frequency-domain perspective. To comprehensively characterize tumor regions, we develop a Frequency Domain Decomposition (FDD) module that separates MRI images into low-frequency components, capturing smooth tumor contours and high-frequency components, highlighting detailed textures and directional edges. To further enhance sensitivity to tumor boundaries, we introduce an Adaptive Laplacian Convolution (ALC) module that adaptively emphasizes critical high-frequency details using dynamically updated convolution kernels. To effectively fuse tumor features across multiple scales, we design a Frequency Domain Cross-Attention (FDCA) integrating semantic, positional, and slice-specific information. We further validate and interpret frequency-domain improvements through visualization, theoretical reasoning, and experimental analyses. Extensive experiments on four public datasets demonstrate that HFF-Net achieves an average relative improvement of 4.48\% (ranging from 2.39\% to 7.72\%) in the mean Dice scores across the three major subregions, and an average relative improvement of 7.33\% (ranging from 5.96\% to 8.64\%) in the segmentation of contrast-enhancing tumor regions, while maintaining favorable computational efficiency and clinical applicability. Our code is available at: {\hypersetup{urlcolor=black}\href{https://github.com/VinyehShaw/HFF.git}{https://github.com/VinyehShaw/HFF}}.

\end{abstract}

\begin{IEEEkeywords}
Brain tumor segmentation, Frequency domain, Multi-modal feature fusion.
\end{IEEEkeywords}

\section{Introduction}

\label{sec:introduction}

\begin{figure}[!t]
    \centering{\includegraphics[width=\columnwidth]{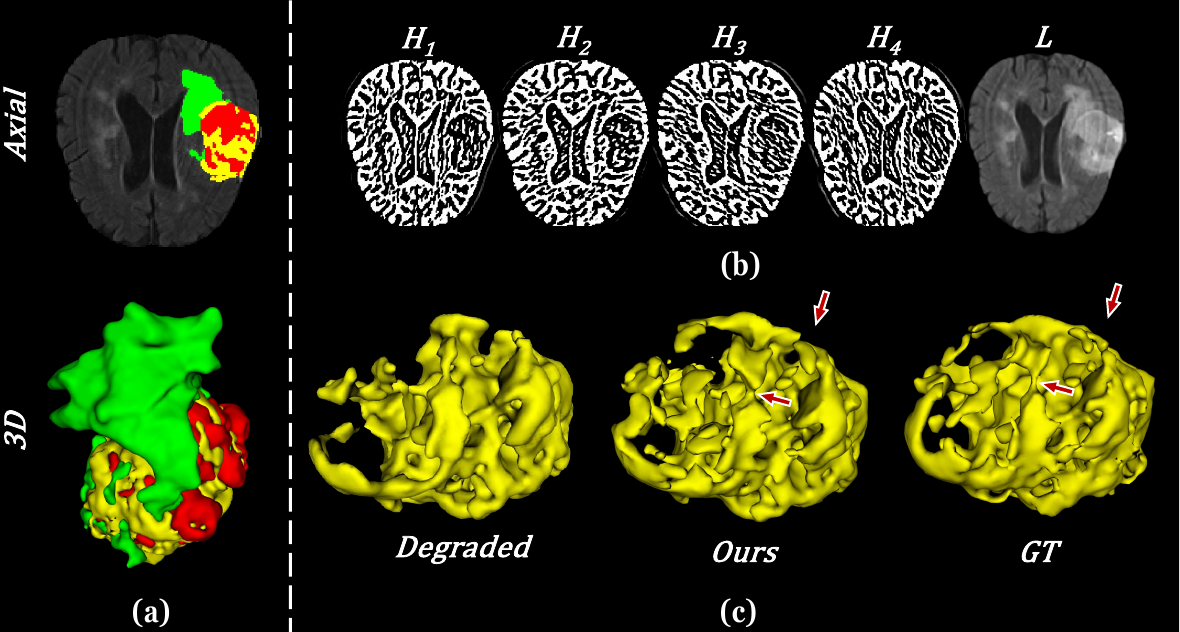}}

    \caption{(a) Segmentation of all tumor regions in a complex brain glioma case. (b) Multi-directional HF (\( H_\textit{i} \)) and LF (\( L\)) input (FLAIR) example in our proposed method in this case. (c) Comparison of the previous work's degraded segmentation performance in contrast-enhancing tumor region with our approach in this case. \textcolor[HTML]{FE0000}{Red} arrows show where our predictions closely match the ground truth.}
    \label{Fig:Fig1}
    \vspace{-5pt}

\end{figure}
\IEEEPARstart{B} {rain} tumors, particularly gliomas, are among the deadliest forms of cancer, with glioblastoma patients having a median survival of only 10 to 14 months \cite{van2010exciting}. Early and accurate diagnosis is essential for effective treatment planning and improved patient outcomes \cite{menze2014multimodal}. Magnetic Resonance Imaging (MRI) serves as the primary non-invasive tool for detecting and characterizing brain tumors, utilizing various sequences such as T1, T1 with gadolinium enhancement (T1Gd), T2, and T2 Fluid Attenuated Inversion Recovery (T2-FLAIR). However, manual delineation of tumor regions in MRI scans is labor-intensive, time-consuming, and subject to inter-observer variability \cite{pereira2016brain}, highlighting the need for reliable and efficient automated segmentation methods. In particular, precise segmentation of contrast-enhancing tumor (ET) regions is clinically critical, as these regions reflect tumor aggressiveness and are strongly associated with patient prognosis \cite{2018对比增强重要的证据}. Their size and morphology directly influence treatment strategies, including surgical resection and radiotherapy planning. Moreover, longitudinal monitoring of ET regions is essential for evaluating treatment response and disease progression \cite{wang2015patterns}.

In conventional clinical workflows for brain tumor diagnosis, texture analysis has been widely applied to quantitatively assess tumor heterogeneity and structural complexity from MRI scans. To further enhance the extraction of diagnostic information, frequency-domain decomposition methods have been incorporated into texture analysis studies \cite{li2018mri,kunimatsu2022texture,kassner2010texture}. These methods can identify subtle texture patterns and signal variations across different spatial scales and frequencies that are typically imperceptible to human observers \cite{zacharaki2009classification, csutak2020differentiating}. Such frequency-derived features have been shown to improve the accuracy and objectivity of tumor grading, subtype classification, and molecular biomarker prediction \cite{kaur2017quantitative}. Importantly, prior studies have demonstrated that these features effectively capture distinctive texture differences between pathological and normal tissues \cite{2010纹理分析所说的纹理特征特点2,davnall2012assessment,soni2019texture}, particularly in contrast-enhancing tumor regions where textural variations are more pronounced \cite{2018增强区域有纹理性,skogen2016diagnostic,yu2021characterizing}.

Despite the demonstrated clinical value of frequency-based texture analysis in brain tumor imaging, most earlier segmentation networks remain confined to spatial-domain representations, overlooking intrinsic texture and directional cues that reflect lesion heterogeneity in medical imaging \cite{litjens2017survey}. As a result, they often exhibit reduced feature stability under imaging artifacts, intensity inconsistencies, or subtle textural variations, which are frequently encountered in clinical settings \cite{yang2020fda}. Such limitations are particularly detrimental to the segmentation of contrast-enhancing brain tumor regions, for which precise boundary delineation is critical for treatment planning and prognosis assessment \cite{2023btsurvey, mazzara2004brain,ellingson2017baseline}. From a modeling perspective, semantic segmentation is essentially a structured feature reconstruction task that learns spatial and semantic correspondence between input images and target masks. Its success hinges on the representational quality of early-stage features, particularly their spatial coherence and semantic discriminability. The omission of frequency-domain cues thus constrains feature expressiveness and limits the network’s capacity to delineate fine-grained tumor boundaries \cite{hajiabadi2021comparison,qiong2025medical}.

Recent research in medical image segmentation have increasingly recognized the advantages of frequency-domain representations in enhancing feature robustness, capturing fine-grained textures, and improving boundary delineation. Discrete wavelet transforms (DWT) have been incorporated to preserve multiscale structural details \cite{singh2022prior}, while dual-tree complex wavelet transforms (DTCWT) have been employed for their stability and directional sensitivity under imaging perturbations \cite{peng2024spectral, garia2025dual}. Directional multiscale transforms such as contourlets have also been explored to better capture anisotropic texture patterns characteristic of tumor heterogeneity \cite{darooei2023optimal}.
Nevertheless, much of this progress treats frequency information as a uniform feature space without fully distinguishing between components that predominantly encode global morphology and those that capture fine-grained textures and directional variations \cite{peng2024spectral, darooei2023optimal, agnes2024wavelet}. Furthermore, frequency decompositions are often incorporated in a shallow, slice-wise manner \cite{zhao2023wranet}, restricting the ability to model volumetric coherence and to exploit the rich multi-scale, multi-modal information inherent in MRI scans. To effectively address the unique challenges of brain tumor segmentation, particularly in accurately delineating contrast-enhancing tumor (ET) regions from multi-modal 3D MRI data, a more systematic integration of frequency-domain features is needed. An ideal framework should not only capture complementary low- and high-frequency information but also preserve volumetric consistency and adapt to the heterogeneous textural and morphological characteristics of tumor regions. 

Motivated by these considerations, we propose HFF-Net, a frequency-aware segmentation framework that rethinks brain tumor analysis from the frequency-domain perspective and explicitly targets accurate delineation of ET regions in multi-modal 3D MRI. It integrates Dual-Tree Complex Wavelet Transform (DTCWT) and Nonsubsampled Contourlet Transform (NSCT) to extract complementary low- and high-frequency features, enhancing shift invariance, noise robustness, and directional sensitivity \cite{lu2018双树复小波, wang2021NSCT优势}. An Adaptive Laplacian Convolution (ALC) module further refines edge and texture representations via dynamic reweighting, while Frequency Domain Cross Attention (FDCA) modules improve global spatial coherence and multi-scale frequency interaction. Training is jointly driven by label-guided supervision and unsupervised dual-branch alignment, implemented via a 3D Dynamic Focal Loss (DFL) that harmonizes low- and high-frequency representations.

To the best of our knowledge, this is the first study to integrate two distinct frequency domain transformation methods for 3D brain tumor segmentation. Our approach harmonizes both high-frequency (HF) and low-frequency (LF) feature attributes through a series of frequency domain techniques to align with the pathological structures characteristic of tumor tissues in MRI images. The key contributions of this study are as follows:\begin{itemize}    
    \item 
An ALC layer is proposed to facilitate the selective adaptation of multi-directional HF features. Employing a Laplace convolution kernel with HF filtering capabilities enhances the model's perception of tissue edge information and texture patterns from MRI images.
    \item 
The FDCA is designed for the processing of anisotropic volumetric MRI by focusing on multidimensional attention extraction within the frequency domain. This module effectively simulates inter-slice variability and can be seamlessly integrated into the encoding process, enabling comprehensive capture and fusion of multi-scale features.

 \item We proposed a novel loss function that incorporates 3D DFL in its unsupervised part to dynamically adjust dual-branch outputs, strategically fostering a harmonized consolidation of HF and LF features for superior segmentation performance.
 \item HHF-Net, a dual-branch network processing and integrating HF and LF features from distinct frequency domain decompositions, demonstrates superior performance in extensive experiments, particularly in mitigating segmentation degradation in enhancing tumor areas critical for clinical brain tumor treatment.
\end{itemize}

\section{Related Works}

\label{Sec:Sec2}

\subsection{Medical Semantic Segmentation}

Semantic segmentation of medical images is critical in modern diagnostics, as it aims to precisely delineate anatomical structures and identify pathological features. Alongside other downstream tasks such as classification and detection, it has witnessed rapid methodological advances, driven by the evolution of deep learning frameworks~\cite{badjie2022deep,kurdi2023brain,khan2023multimodal,maqsood2021efficient,rajinikanth2022glioma, zhai2025dsleepnet, li2024sid,zhang2024depth,10323083,wan2024sentinel,miao2025rethinking,miao2025laser,duan2023dynamic, duan2025parameter, li2025unified, li2025bp, chang2023design, chen2024hint}.

\textbf{Single-Modality Methods.} Recent advancements in single-modality segmentation leverage CNNs and their variants. The seminal UNet architecture \cite{3DUNET} and its successors, nnUNet \cite{2021nnunet} and UNETR \cite{2022unetr}, enhance feature extraction through layered encoder-decoder paths. Attention mechanisms, as in Attention UNet \cite{2018attentionunet}, further refine models' abilities to focus on relevant features. ResUNet 3D \cite{zhang2018road} and paired attention in UNETR++ \cite{shaker2024unetr++} push boundaries with detailed spatial focus. Other notable methods include the Two-stage Cascaded U-Net \cite{jiang2020two} and SGEResU-Net \cite{liu2022sgeresu} combining residual learning and group-wise attention for improved brain tumor segmentation.

However, these single-modality approaches inherently suffer from limited contextual understanding, as they rely on information from a single imaging sequence. This constraint reduces their ability to capture the diverse pathological characteristics of brain tumors across modalities, limiting their effectiveness in complex clinical scenarios \cite{zhao2022modality}.

\textbf{Multi-Modality Methods.} Multi-modality methods integrate data from various imaging techniques to address complex diseases comprehensively. The BRATS benchmark \cite{menze2014multimodal} emphasizes using multimodal data, combining MRI, Computed Tomography (CT), and Positron Emission tomography (PET) scans. TransUNet \cite{chen2021transunet} combines Vision Transformer (ViT) \cite{2020VIT} with UNet for enhanced feature integration. H$^2$NF-Net \cite{jia2021hnf} leverages multi-modal MR imaging for enhanced brain tumor segmentation. SA-LuT-Nets \cite{yu2021sa} employs sample-adaptive intensity lookup tables to improve segmentation performance in multimodal datasets.

While most multi-modality methods improve segmentation by integrating information from multiple sequences, but they often neglect frequency-specific structural and textural characteristics. This oversight limits their effectiveness in accurately delineating complex regions \cite{qiao2024medical}. Some methods compromise practicality in clinical workflows due to increased training and deployment complexity. For example, \cite{jiang2020two} introduces a two-stage cascade to address coarse boundary segmentation, but its multi-stage complexity limits clinical applicability.

In summary, although existing methods have advanced brain tumor segmentation, they inadequately address the intrinsic textural heterogeneity and directional complexity of tumor tissues \cite{ahamed2023review}, particularly in contrast-enhancing regions \cite{liu2024innovative}. In order to address the challenges of multi-modal feature heterogeneity and directional texture complexity, we explicitly design the Frequency Domain Cross Attention (FDCA) and Adaptive Laplacian Convolution (ALC) layer to operate on early-stage frequency decompositions obtained from the Frequency Domain Decomposition (FDD) module, enhancing spatial coherence and fine-grained texture discrimination across modalities. Our method draws inspiration from clinical texture analysis and explicitly considers the structural complexity and modality complementarity of brain MRI, addressing these limitations through a frequency-domain dual-branch design. This design fully exploits complementary information across modalities while maintaining an end-to-end framework that balances segmentation performance and clinical applicability.

\subsection{Wavelet and Contourlet-Based Deep Neural Networks}
Frequency-domain transforms have been extensively utilized across a wide range of computer vision tasks, including representation learning  \cite{
xu2020learning,li2023discrete,liu2023improving2,zhu2024wavelet},  image generation \cite{jiang2021focal,phung2023wavelet}, and image super-resolution \cite{fuoli2021fourier,liu2023spectral}. These techniques offer the ability to disentangle structural information across different spatial scales and frequency bands, thereby enhancing model capacity in capturing both global and local patterns. 

In the domain of medical image segmentation, frequency-domain transforms, particularly wavelet-based techniques, have been widely incorporated into deep neural networks (DNNs) due to their effectiveness in multi-scale feature representation and noise reduction. A prevalent strategy involves embedding wavelet transforms in specific components of segmentation networks, such as preprocessing, postprocessing, or replacing downsampling layers \cite{8641484, zhou2023xnet}. For example, Azimi et al. \cite{2018aerial小波} proposed a wavelet-enhanced FCN to mitigate detail loss in small object segmentation. Lu et al. \cite{lu2018双树复小波} employed the Dual-Tree Complex Wavelet Transform (DTCWT) to improve CNNs' ability to capture structural textures and suppress noise in medical images. Recent studies further embedded Haar wavelet transforms within U-Net++ architectures to enhance feature extraction for complex pulmonary nodule segmentation \cite{agnes2024wavelet}. 

Beyond wavelets, the contourlet transform has emerged as a more flexible and powerful alternative. For example, \cite{ji2025structural} effectively exploited contourlet decomposition in semantic segmentation, demonstrating its superior ability to capture multi-scale and directional structural textures, which are difficult to represent by wavelets. Nevertheless, the classical contourlet transform still suffers from subsampling-induced spatial aliasing and limited localization accuracy, especially in high-frequency regions \cite{lu2006new}. To further address these limitations, the Nonsubsampled Contourlet Transform (NSCT) \cite{pyrexc滤波器nsct} was proposed, which removes the subsampling operations and adopts nonsubsampled filter banks. This design enhances spatial localization and directional selectivity while better preserving fine-grained high-frequency textures, making it particularly suitable for dense prediction tasks that require detailed texture characterization \cite{2016轮廓波1}. By providing multi-directional decomposition without downsampling, NSCT effectively captures anisotropic textures and fine-grained edge information, which is particularly beneficial for segmenting complex anatomical structures in medical images \cite{2016轮廓波3}. In brain MRI segmentation, Reddy et al. \cite{2017brain轮廓波} integrated NSCT with active contour models to improve the delineation of weak, blurred, and irregular tumor boundaries.

\begin{figure*}[t]
  \centering
   \includegraphics[width=1\linewidth]{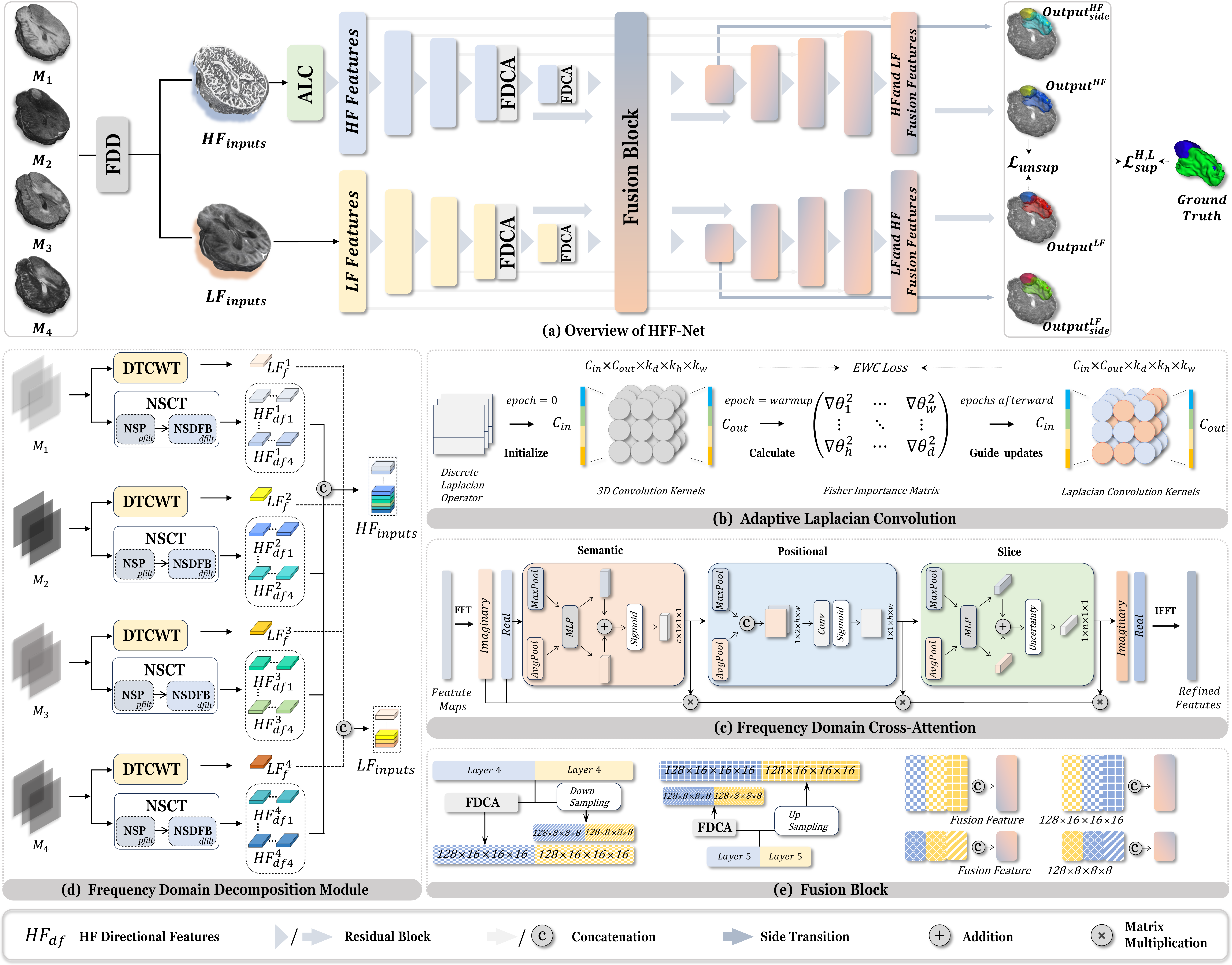}

   \caption{(a) Architecture of our HFF-Net: A multimodal dual-branch network decomposing and integrating multi-directional HF and LF MRI features with three components: ALC, FDCA, and FDD. It uses $\mathcal{L}_{\textit{unsup}}$ for output consistency between branches and $\mathcal{L}_{\textit{sup}}^\textit{H,L}$ to align each branch's main and side outputs with ground truth. (b) Our ALC uses elastic weight consolidation to dynamically update weights, maintaining HF filtering functionality while extracting features from multimodal and multi-directional inputs. (c) FDCA enhances the extraction and processing of anisotropic volumetric features in MRI images through multi-dimensional cross-attention mechanisms in the frequency domain. (d) FDD processes multi-sequence MRI slices by decomposing them into HF and LF inputs using distinct frequency domain transforms. (e) The fusion block integrates the deep HF and LF features from the deep layers during the encoding process.}

   \label{fig:fig2}
\end{figure*}
Despite the demonstrated effectiveness of wavelet- and contourlet-based deep networks, prior methods predominantly adopt a fragmented design, applying a single type of frequency transform in isolated modules without fully leveraging frequency-domain information across the entire network. This limits their ability to jointly capture complementary low- and high-frequency features essential for precise tumor delineation \cite{hajiabadi2021comparison}. In order to overcome the fragmented use of frequency information in previous methods, we introduce a unified FDD module in HFF-Net, which systematically integrates two clinically validated
frequency-domain transforms for comprehensive extraction of complementary low- and high-frequency features critical to precise brain tumor segmentation. DTCWT is chosen for its shift invariance and noise robustness in extracting smooth anatomical contours, while NSCT provides superior directional selectivity and localization of fine-grained textures. This specific combination is not arbitrary but motivated by both prior segmentation studies that successfully embedded these transforms and established clinical practices in brain tumor texture analysis, where frequency decomposition has been proven effective for revealing subtle pathological patterns. Together, these two transforms enable HFF-Net to comprehensively characterize the structural and textural Heterogeneity of brain tumors, particularly in clinically critical contrast-enhancing regions. 

\section {Proposed Method}
\label{sec:sec3}
\subsection{Overview}
As illustrated in \Cref{fig:fig2} (a), HFF-Net integrates three key modules: Frequency Domain Decomposition (FDD, detailed in \ref{subsec:D}), Adaptive Laplacian Convolution (ALC, detailed in \ref{subsec:B}), and Frequency Domain Cross-Attention (FDCA, detailed in \ref{subsec：C}).

Initially, the FDD module decomposes multimodal MRI inputs into low-frequency (LF) and multi-directional high-frequency (HF) sub-bands, producing frequency-specific features \(x^L \in \mathbb{R}^{M \times D \times H \times W}\) and \(x^H \in \mathbb{R}^{4M \times D \times H \times W}\), where \( M \) is the number of MRI modalities, and each HF component comprises four directional sub-bands. To address the modality diversity and directional complexity of these HF features, the subsequent ALC layer adaptively enhances discriminative responses using Laplacian-initialized convolutions weighted by Fisher information. refined HF features and LF features are then separately processed through dedicated LF and HF encoders consisting of residual-connected layers. At deeper encoder layers, the FDCA module employs frequency-domain attention to integrate LF and HF representations, enhancing cross-frequency consistency and modality complementarity for subsequent fusion and decoding.

Encoded features are then fused (\Cref{fig:fig2} (e)) and expanded via symmetric decoder layers, incorporating skip connections between corresponding encoder-decoder layers and intermediate side-output predictions to facilitate supervised learning. HFF-Net is optimized through joint supervised segmentation losses (primary and side outputs) and an unsupervised dual-output consistency loss via 3D Dynamic Focal Loss (DFL), ensuring robust integration and mutual enhancement between LF and HF branches. The final segmentation result \(y \in \mathbb{R}^{C \times D \times H \times W}\) is selected from the optimal primary branch prediction, where \( C \) is the number of segmentation classes (e.g., \( C \) = 2 or 4 for brain tumor segmentation).

\subsection{Adaptive Laplacian Convolution}
\label{subsec:B}
The Adaptive Laplacian Convolution (ALC) layer, as illustrated in \Cref{fig:fig2} (b), is designed specifically to manage the complexity arising from multi-directional high-frequency (HF) sub-bands generated by the Frequency Domain Decomposition (FDD) module. Considering the substantial dimensionality introduced by multiple MRI modalities, each decomposed into four directional sub-bands, manual feature selection becomes impractical. Hence, the ALC module autonomously selects and enhances informative directional and modal HF features, adaptively emphasizing structural details essential for accurate segmentation. The enhanced HF representations produced by ALC subsequently provide robust inputs for further integration in the FDCA module.

Inspired by the discrete Laplace operator's ability to accentuate rapid intensity changes at edges through its sensitivity to local signal variations via second derivative calculations, and its isotropic nature that ensures uniform detection of edges and details in all directions \cite{gonzalez2017edge}, we have integrated it into our convolution kernel within the ALC layer. This innovative technical combination enables the model to selectively emphasize high-frequency features from multimodal anisotropic data, thereby enhancing the detection and interpretation of complex brain tumor boundaries and texture details, thus further elevating the accuracy and relevance of segmentation outputs.

In the domain of continual learning,  Elastic Weight Consolidation (EWC) is a crucial technique that addresses the issue of catastrophic forgetting in neural networks, which occurs when networks are trained on new tasks \cite{wang2024comprehensive}. The core principle of EWC is to protect important weights from being overly modified by new tasks during model training by adding a regularization term. Specifically, the importance of each weight is usually estimated by calculating the second-order derivative of the loss function, known as the Fisher information matrix (FIM) \cite{wang2024comprehensive}. When learning new tasks, the original loss function is augmented with this regularization term, allowing the model to retain as much memory of previous tasks as possible while it learns new ones. This method effectively mitigates catastrophic forgetting by striking a balance between retaining old knowledge and acquiring new information.
Inspired by this, the ALC layer employs a similar strategy to achieve a balance between maintaining crucial weights for HF filtering and updating weights for adaptive learning of HF feature importance.  Initially, the convolutional layer's kernel weights \( k \in \mathbb{R}^{C_{\text{out}} \times C_{\text{in}} \times H_d \times W_d \times D_d} \) are initialized with a discrete Laplace operator, where \( C_{\text{out}} \) represents the number of output channels, \( C_{\text{in}} \) denotes the number of input channels, and the dimensions \( H_d \), \( W_d \), and \( D_d \) specify the height, width, and depth of the discrete Laplace operator, respectively. A benchmark kernel of identical dimensions is established for comparison. During training, \( \mathcal{L}_{\text{ewc}} \) is continuously applied to the weights in alignment with a benchmark Laplacian kernel of identical dimensions. This regularization is specifically designed to guide the update process, ensuring that crucial weights are not excessively modified, thereby preserving their characteristics in close resemblance to the original discrete Laplace operator. As the training reaches a certain epoch, the importance of each weight within the kernel is assessed using the FIM. This process can be expressed as:
\begin{equation}
F_p = \frac{1}{N} \sum_{i=1}^N \left(\nabla_p \mathcal{L}(x_i, y_i; \theta)\right)^2,
\end{equation}
where \(F_p\) represents the FIM for the parameters \(p\) of the convolutional kernel in the ALC layer, \(N\) is the total number of data batches, \(\nabla_p \mathcal{L}(x_i, y_i; \theta)\) represents the gradient of the loss function \(\mathcal{L}\) with respect to the convolutional kernel parameters \(p\) in the ALC layer, \(x_i\) and \(y_i\) are the input to the model and the corresponding ground truth output of the \(i\)-th data batch, respectively, and \(\theta\) specifically denotes the weights of the convolutional kernel in the ALC layer. Then Z-score thresholding is employed to distinguish between important and less important weights:
\begin{equation}
W_{\textit{imp}} = \{p \mid F_p > \mu_F + k \cdot \sigma_F\},
\end{equation}
where \(W_{\textit{imp}}\) represents the set of important weights, \(F_p\) is the Fisher information for each parameter \(p\), \(\mu_F\) is the mean of the FIM across all parameters, \(\sigma_F\) is the standard deviation of the FIM, and \(k\) is a constant factor that scales the standard deviation to set the threshold. Subsequently, important weights are frozen to maintain the kernel’s HF filter capabilities, while less critical weights are dynamic, allowing for ongoing updates and adjustments. 

In this study, the epoch for calculating weight importance is set after the warmup phase, as gradients become more consistent and stable following this training period.

\subsection{Frequency Domain Cross-Attention}
\label{subsec：C}
As illustrated in \Cref{fig:fig2} (c), the Frequency Domain Cross-Attention (FDCA) module refines and integrates deeper features extracted by the preceding ALC and LF encoding layers, enhancing anisotropic feature representation through cross-attention mechanisms. The module operates on multi-scale feature maps \( F \in \mathbb{R}^{c \times n \times h \times w} \) sourced from various depths within the network encoder. Here, \( c \) denotes the number of channels, \( n \) represents the number of slices within the volumetric feature, and \( h \) and \( w \) represent the height and width of each slice, respectively.  
 FDCA employs semantic, positional, and slice attention, starting with decomposing the input feature map the \( F \) into the real \( {F_\text{r}} \) and imaginary \( F_{\text{i}} \) components via the Fast Fourier Transform (FFT). This enables efficient attention extraction from deep features during training, enhancing feature distinction while maintaining low computational costs. Semantic \( M_s \)\( \in \mathbb{R}^{c \times 1 \times 1 \times 1} \), positional \( M_p \)\( \in \mathbb{R}^{1 \times 1 \times h \times w} \), and slice \( M_n \)\( \in \mathbb{R}^{1 \times n \times 1 \times 1} \) attention maps are computed sequentially. These attention maps operate on the real part of the feature map through broadcasted multiplication \( \otimes \):
\begin{equation}
\begin{aligned}
F_{r1} &= M_s(F_r) \otimes F_r, &\quad F_{i1} &= M_s(F_r) \otimes F_i,\\
F_{r2} &= M_p(F_{r1}) \otimes F_{r1},&\quad F_{i2} &= M_p(F_{r1}) \otimes F_{i1},\\
F_{r}' &= M_n(F_{r2}) \otimes F_{r2},&\quad F_{i}' &= M_n(F_{r2}) \otimes F_{i2}.
\end{aligned}
\end{equation}
Finally, the attention-modulated features in the frequency domain are transformed back into the spatial domain via the Inverse Fast Fourier Transform (IFFT), which consolidates complementary frequency-specific details from HF and LF branches, preparing unified and discriminative feature representations for the subsequent decoding phase.

Semantic attention refines channel features \( c \) by applying maximum \( P_{\text{max}} \) and average \( P_{\text{mean}} \) pooling operations to the real-valued frequency domain features \( F_r \). These operations distill essential multimodal and multidirectional information, which is then encoded via affine transformations \( W_1 \) and \( W_2 \), respectively. The combined outputs are passed through a sigmoid function to produce semantic attention weights \( M_s \):
\begin{equation}
M_s(F_r) = \sigma(W_1 \cdot P_{\text{max}}(F_r) + W_2 \cdot P_{\text{mean}}(F_r)).
\end{equation}

Positional attention extracts crucial tissue contours and structural information from all slices, regardless of the \( h \) and \( w \) dimensions of the volumetric data. Maximum pooling \( P_{\text{max}} \) and average pooling \( P_{\text{mean}} \) are applied to the semantically enhanced frequency domain features  \( F_{r1} \), generating a dual-channel positional feature map refined by a 3D convolutional layer. Normalization using a sigmoid function produces positional attention weights \( M_p \):
\begin{equation}
M_p(F_{r1}) = \sigma(\textit{Conv}(\textit{Concat}(P_{\text{max}}(F_{r1}), P_{\text{mean}}(F_{r1})))).
\end{equation}

Slice attention enhances the network's understanding of key features along the slice dimension \(n\) by integrating statistical information and managing uncertainty. Important statistics are extracted using maximum \(P_{\text{max}}\) and average \(P_{\text{mean}}\) pooling operations, combined via linear transformations:
\begin{equation}
S = W_1 \cdot P_{\text{max}}(F_{r2}) + W_2 \cdot P_{\text{mean}}(F_{r2}).
\end{equation}
We employ learnable linear transformations \(W_\mu\), \(W_r\), and \(W_d\) to derive the parameters of the mean vector \(\mu\), the low-rank covariance factor \(R\), and the diagonal covariance matrix \(D\). Note, to ensure the positiveness of the covariance matrix, an exponential function is applied to the derived matrix \(D\):
\begin{equation}
\mu = W_\mu(S), \quad R = W_r(S), \quad D = \exp(W_d(S)).
\end{equation}
These parameters contribute to modeling a low-rank Gaussian distribution \(\mathcal{N}_{L}(\mu, \Sigma)\), where \(\Sigma = RR^T + D\). The slice attention weights \(M_n\) are derived by sampling a vector \(v\) from this distribution and processing it through a sigmoid function:
\begin{equation}
M_n = \sigma(v), \quad v \sim \mathcal{N}_{L}(\mu, \Sigma).
\end{equation}
\begin{figure}[t]
  \centering   \includegraphics[width=\linewidth]{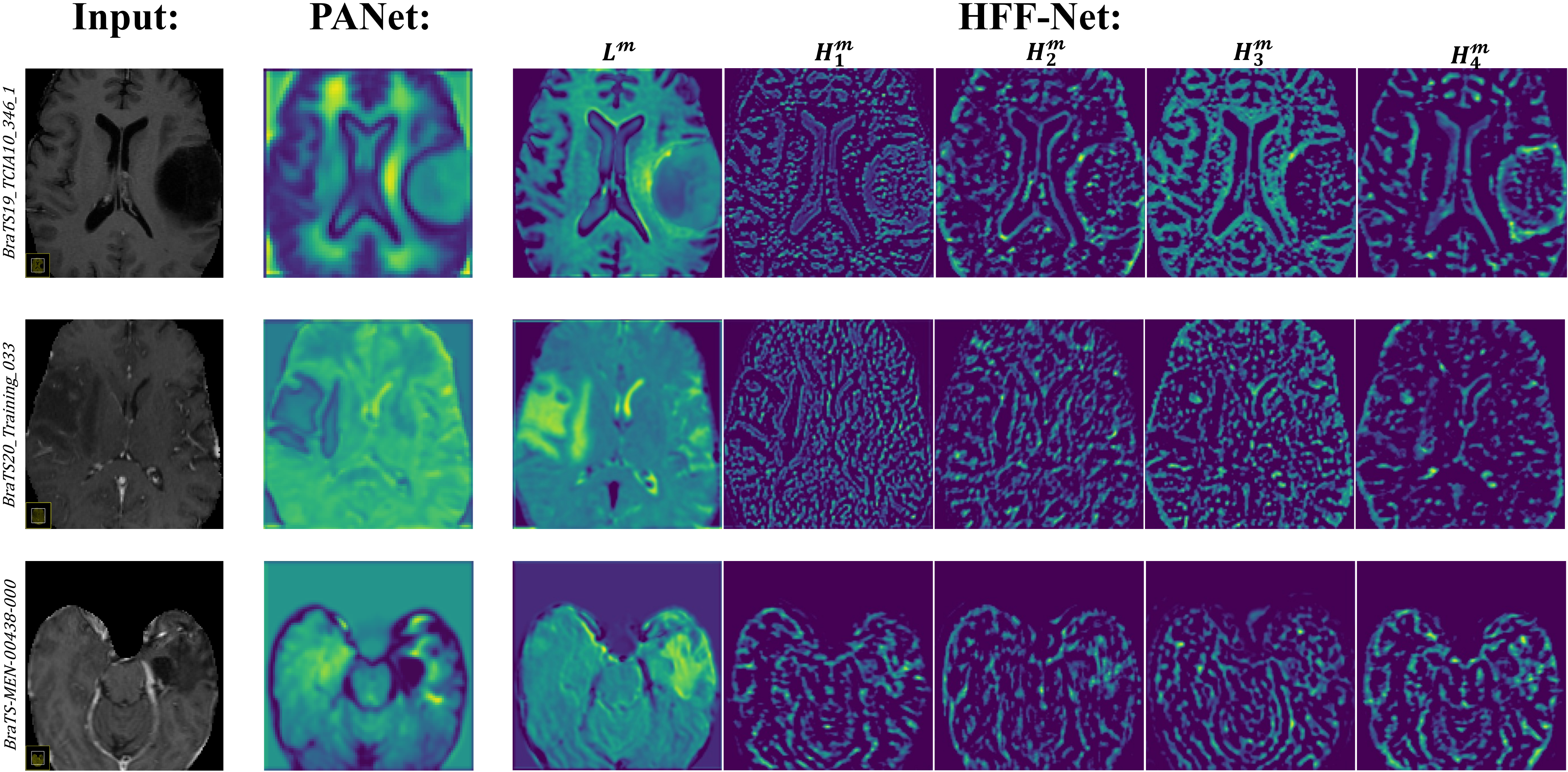}
   \caption{Feature maps between PANet \cite{PANet} and HFF-Net. Conventional methods often overlook critical frequency-specific cues for brain tumor segmentation. In contrast, HFF-Net decouples frequency components to preserve low-frequency structural integrity while multi-directionally enhancing high-frequency tumor boundary details and texture granularity.}
   \label{fig:fig3}  
\end{figure} 
\vspace{-20pt}
\subsection{Frequency Domain Decomposition Module}
\label{subsec:D}
As depicted in \Cref{fig:fig2} (d), the Frequency Domain Decomposition (FDD) module serves as the initial step of the segmentation framework by transforming multi-sequence MRI scans into distinct low-frequency (LF) and high-frequency (HF) sub-bands, while preserving frequency-specific characteristics and spatial correspondence. These sub-bands are then fed into corresponding encoders, with HF inputs further refined by the Adaptive Laplacian Convolution (ALC) module to enhance directional feature representation. \Cref{fig:fig3} further compares the contrast in feature map representation between conventional approaches (e.g., PANet) and HFF-Net on three dataset cases from the T1c modality, the most discriminative modality for enhancing tumor segmentation. While PANet integrates multi-frequency information in an undifferentiated manner, potentially blurring structural fidelity, HFF-Net explicitly disentangles frequency components via the FDD module, facilitating complementary encoding of coarse and fine-grained features for enhanced segmentation precision.

The LF features are obtained using the Dual-Tree Complex Wavelet Transform (DTCWT), which primarily captures large-scale structural information such as the overall shape and contour of tumors and surrounding tissues. This process can be formulated as:
\begin{equation}
f_m(x, y) \rightarrow \text{DTCWT}_{\textit{r}, \textit{i}} \rightarrow L_m^{S} \text{,}
\end{equation}
where \(L_m^{S}\) denotes the LF subband components extracted from the \(m\)-th modality of the MRI image \(f(x, y)\) , incorporating both real (\(\textit{r}\)) and imaginary (\(\textit{i}\)) parts. The number of decomposition levels is set to one, balancing essential LF contour information and computational efficiency. To highlight the advantages of DTCWT's LF decomposition, we first recall the formulation of the traditional discrete wavelet transform (DWT):
\begin{equation}
L_{\mathrm{DWT}}(f) = \sum_{k} \langle f, \phi_{j,k} \rangle \, \phi_{j,k}(x, y),
\end{equation}
where \(L_{\mathrm{DWT}}(f)\) denotes the low-frequency component of the input image \(f(x,y)\), such as a 2D brain MRI slice. Here, \(\phi_{j,k}(x, y)\) is the scaling function at scale \(j\) and spatial location \(k\), and \(\langle f, \phi_{j,k} \rangle\) is the inner product between the image and the scaling function, which is the LF coefficient at that scale and position. Note that the inner product is defined over the square-integrable function space \(L^2(\mathbb{R}^2)\) as:
\begin{equation}
\langle f, \phi_{j,k} \rangle = \iint_{\mathbb{R}^2} f(x, y) \, \phi_{j,k}(x, y) \, dx \, dy.
\end{equation}
The DTCWT, in comparison, performs LF decomposition via:
\begin{equation}
L_{\mathrm{DTCWT}}(f) = \sum_{k} \left( \langle f, \phi_{j,k}^{r} \rangle + i \langle f, \phi_{j,k}^{i} \rangle \right) \, \phi_{j,k}^{r}(x, y),
\end{equation}
where \(\phi_{j,k}^{r}(x, y)\) and \(\phi_{j,k}^{i}(x, y)\) form a pair of Hilbert transform scaling functions representing the real and imaginary components of the complex wavelet, respectively. This Hilbert pair is designed to approximately satisfy the complementary energy condition:
\begin{equation}
|\hat{\phi}^{r}(\omega)|^2 + |\hat{\phi}^{i}(\omega)|^2 \approx 1, \quad \forall \, \omega \in [-\pi, \pi),
\end{equation} where \(\omega\) denote the angular frequency.
This structural design ensures that the real and imaginary parts complement each other in the frequency domain, thereby preserving stable responses across all frequencies. This stability is crucial for reliably capturing signal variations. In the context of MRI, it may correspond to subtle misalignments caused by patient motion or intensity inconsistencies (e.g., inter-slice brightness variations) resulting from physiological dynamics like breathing or pulsation.

Whether using traditional DWT or NSCT, their LF components are generated by real-valued and typically direction-insensitive scaling functions. These lack the ability to represent phase and directional information, making them more sensitive to small shifts or subtle structural changes in the image, potentially degrading the stability and robustness of extracted features. Therefore, the DTCWT is employed for LF decomposition, as its complex representation offers improved approximate shift invariance compared to other transforms.

For HF features, which capture sharp edges and
minute anatomical variations, we employ the NSCT to extract complex multi-directional textures and edge information from brain tumor images, the process is defined as: \begin{equation}
f_m(x, y) \rightarrow \text{NSP}_{\text{pfilt}} \rightarrow \text{NSDFB}_{\text{dfilt}} \rightarrow H_{m,d}^{\textit{S}} \text{.}
\end{equation}
Here, the Nonsubsampled Pyramid $\text{NSP}_{\text{pfilt}}$ is implemented using the \textquotesingle{}\textit{pyrexc}\textquotesingle{} filter bank derived from one-dimensional filters with two vanishing moments \cite{pyrexc滤波器nsct}. The Nonsubsampled Directional Filter Bank $\text{NSDFB}_{\text{dfilt}}$, based on the \textquotesingle{}\textit{cd}\textquotesingle{} filter bank proposed by Cohen and Daubechies \cite{cd滤波器}, enables arbitrary directional decomposition, in contrast to the DTCWT, which is limited to six fixed orientations. This provides greater flexibility for adapting to various anatomical edge structures. $H_{m,d}^{\textit{S}}$ represents the HF subbands extracted at direction $d$ for each modality, with two levels for both pyramid and directional decomposition.

To illustrate the potential aliasing artifacts introduced by downsampling in high-frequency decomposition, we examine the frequency-domain formulation of the DTCWT, defined as:

\begin{equation}\small
\hat{H}(\omega_x, \omega_y) = \sum_{k,l=0}^{1} \hat{g}(d\omega_x + k\pi, d\omega_y + l\pi) \cdot \hat{\psi}^{r,i}(d\omega_x, d\omega_y),
\end{equation}
where \(\hat{g}(\cdot)\) denotes the frequency-domain representation of the HF component, and \(d = 2\) is the downsampling factor. The indices \(k, l \in \{0, 1\}\) represent spectral shifts in frequency caused by the subsampling process. This formulation reveals that the downsampling process duplicates frequency components and introduces spectral overlapping, which may distort edge and texture structures, issues that are inherently avoided in the nonsubsampled NSCT framework.

In summary, our FDD module integrates DTCWT and NSCT to separately extract structural and textural cues through low- and high-frequency decomposition. DTCWT provides robust low-frequency features via shift invariance and complex phase response, while NSCT captures high-frequency details with aliasing-free, directionally selective analysis. Aligned with the structural heterogeneity and texture patterns of brain MRI, this design improves the spatial coherence and semantic clarity of early-stage features. Compared to traditional DWT or fixed-orientation transforms, it achieves a better trade-off between spatial stability and directional sensitivity, thereby supporting more accurate segmentation.

\subsection{Loss Function}

The total loss function $\mathcal{L}_{\textit{total}}$ of the model can be summarized as:
\begin{equation}
\mathcal{L}_{\textit{total}} =\mathcal{L}_{\textit{sup}} + \lambda_1 \mathcal{L}_{\textit{unsup}} + \lambda_2 \mathcal{L}_{\textit{ewc}},
\label{eq:total_loss}
\end{equation}
where $\mathcal{L}_{\textit{sup}}$ represents the supervised loss and $\mathcal{L}_{\textit{unsup}}$ denotes the unsupervised loss, aligning the consistency of outputs between the dual branches. The weight $\lambda_1$ dynamically adjusts the weight of $\mathcal{L}_{\textit{unsup}}$ during training. Additionally, $\lambda_2$ adjusts the weight of the regularization term in the ALC layer that maintains the HF filtering properties. Specifically, the supervised loss
$\mathcal{L}_{\textit{sup}}$ comprises the LF supervised loss $\mathcal{L}_{\textit{sup}}^{\textit{F}}(\cdot)$, the HF supervised loss $\mathcal{L}_{\textit{sup}}^{\textit{H}}(\cdot)$, both employing the Dice Loss \cite{diceloss}, and the elastic weight consolidation loss $\mathcal{L}_{\textit{ewc}}(\cdot)$ associated with the ALC layer. The supervised loss can written as:
\begin{equation}
\mathcal{L}_{\textit{sup}} = \sum_{\substack{f \in \{\textit{L}, \textit{ H}\} \\ t \in \{\textit{main}, 
 \textit{ side}\}}} \mathcal{L}_{\textit{sup}}^{f}(\hat{y}^{f_{t}}_{i}, y_i) + \mathcal{L}_{\textit{ewc}}(\theta,D),
\end{equation}
where $\hat{y}_i^f$ denotes the predicted result for the \(i\)-th MRI slice in frequency band \(f\), and \(t\) indicates the type of output from the model. Respectively, \(y_i\) represents the ground truth for the \(i\)-th MRI slice. Additionally, \(\theta\) denotes the parameters of the convolution kernels across all channels in the ALC layer, adjusted by their importance for continued updates, while \(D\) represents the discrete Laplacian matrices of corresponding dimensions.
Similarly, the unsupervised loss $\mathcal{L}_{\text{unsup}}$ employs a 3D DFL that uses the prediction from one branch as a pseudo-label to supervise the other branch and vice versa. The unsupervised loss can be defined as follows:
\begin{equation}
\mathcal{L}_{\textit{unsup}} = \mathcal{L}_{\textit{unsup}}^{\textit{FF}}(\hat{y}^{\textit{L}}, \hat{y}^{\textit{H}}),
\end{equation}
where $\hat{y}^{\textit{L}}_i$ and $\hat{y}^{\textit{H}}_i$ denote the main outputs of the model from the LF and HF branches, respectively. The final segmentation outcome is determined by comparing and selecting the superior main output from the two branches.

\textbf{3D Dynamic Focal Loss.} The 3D DFL dynamically balances discrepancies between the outputs of the two branches, ensuring effective gradient updates, especially in later training stages. This refinement is crucial for the model's performance as it addresses complex discrepancies between the outputs of different branches. Using 3D FFT, the 3D DFL method converts each volumetric data patch from the spatial to the frequency domain: 
\begin{equation}
O_{f}(u, v, w)_{\substack{f \in \{L, H\}}} = R(u, v, w) + I_i(u, v, w),
\end{equation}
where \(O_f\) denotes the frequency domain output for \(f \in \{L, H\}\), \(R\) and \(I_i\) are the real and imaginary parts, respectively, and \(u, v, w\) are the frequency domain coordinates. The Euclidean amplitude distance between the outputs of the two branches is computed as:
\begin{equation}
D(O_L, O_H) = \sum_{u, v, w} \left| O_L(u, v, w) - O_H(u, v, w) \right|^2 .
\end{equation}
Dynamic weights are calculated based on amplitude differences:
\begin{equation}
w(u, v, w) = |O_L(u, v, w) - O_H(u, v, w)|^\alpha,
\end{equation}
where the exponent \( \alpha \) is a predefined parameter that adjusts the sensitivity of the weight, thereby better addressing differences between the two frequency domain outputs. This approach allows the model to dynamically adjust the loss function weights based on amplitude differences between frequency components, effectively guiding the learning process to focus on significant discrepancies in the frequency domain, optimizing segmentation performance. Finally, the total loss for the entire volume is computed as a weighted average of the Euclidean amplitude distance and dynamic weights, defined as:

\begin{equation}\small
\text{DFL} = \frac{1}{N} \sum_{u, v, w} w(u, v, w) \cdot D(O_L(u, v, w), O_H(u, v, w)),
\end{equation}
\normalsize
where \( N \) is the total number of voxels, ensuring that the loss function is normalized over the entire data volume.

\section{Experiment}
\label{sec:sec4}

\subsection{Datasets}


\subsubsection{BraTS2023-MEN} This dataset \cite{labella2023asnr} includes 1650 multiparametric MRI scans of intracranial meningiomas, with 1000 training, 140 validation, and an undisclosed test set. Each MRI series includes T1-weighted pre-contrast (T1N), T1-weighted post-contrast (T1C), T2-weighted (T2W), and T2-weighted FLAIR (T2F). The training set is split into 80\% for training, 15\% for validation, and 5\% for testing.

\subsubsection{MSD BTS Task} This dataset \cite{MSD1} consists of 750 multiparametric MRI scans of glioblastoma and lower-grade glioma, using T1, T1-Gd, T2, and FLAIR sequences. It targets edema, enhancing, and non-enhancing tumor regions. Of the images, 484 are annotated and divided into 80\% training, 15\% validation, and 5\% testing.

\subsubsection{BraTS2020 and BraTS2019} Both datasets \cite{BRA1, BRA2} provide multimodal scans with T1, T1Gd, T2, and T2-FLAIR modalities. The 2020 set has 369 annotated samples, and the 2019 set has 335. We performed five-fold cross-validation on these datasets.
\subsection{Implementation Details}
Our proposed HFF-Net is implemented in PyTorch 2.1.2, CUDA 11.8, and trained, validated, and tested on an NVIDIA GeForce RTX 4090 (24GB) GPU. Experiments are conducted on four publicly available brain tumor segmentation datasets: BraTS2019, BraTS2020, BRATS2023-MEN, and MSD BTS. The network is trained using stochastic gradient descent (SGD) with a momentum of 0.9, a weight decay of $5 \times 10^{-5}$, and an initial learning rate of 0.3, which decays by a factor of 0.53 every 50 epochs. Each model is trained for 350 epochs with a batch size of 1. The regularization weight in the ALC layer is fixed at $5 \times 10^{-6}$, and a warm-up period of 40 epochs is employed to gradually stabilize training. During the warm-up phase, the weight \(\lambda\) for the unsupervised loss increases linearly with the training epoch, following \(\lambda^{(t)} = \lambda_{\text{max}} \cdot \frac{t}{T}\), where $\lambda_{\text{max}}$ is set to 15, \(t\) is the current epoch and \(T\) is the total number of epochs. To enhance training stability, we apply Kaiming normal initialization \cite{he2015delving} to all convolutional layers, standard normal initialization to fully connected layers, and set all bias terms to zero. To prevent overfitting and improve generalization, several data augmentation strategies are employed during training, including random flipping along each anatomical axis, random rotations within $\pm10^{\circ}$, and random cropping centered on brain regions to extract $128 \times 128 \times 128$ voxel patches. All input volumes are standardized using Z-score normalization (zero mean and unit variance). For all experiments involving five-fold cross-validation, we randomly partition each dataset into five approximately equal folds. Stratified sampling is employed where applicable to preserve class distribution. In each fold, the model is re-initialized and trained from scratch on four folds, while the remaining fold is used for validation. All preprocessing and augmentation steps, including Z-score normalization and patch extraction, are applied independently within each training fold to avoid data leakage.

\subsection{Experimental Results}
\begin{table}
    \centering 
    \begin{subtable}
    \centering
    \vspace{-20pt}
    \caption{Comparison of Segmentation Performance on BRATS2023-MEN Dataset. \textcolor[HTML]{FE0000}{Red} Font Denotes the Best Results And \textcolor[HTML]{00009B}{Blue} Font Represents The Second Best Results}
\label{tab:tab1}
\resizebox*{1\linewidth}{0.15\textheight}{
\begin{tabular}{c|cccccccc} 
\hline
{\color[HTML]{000000} }                                  & \multicolumn{4}{c|}{{\color[HTML]{000000} \textbf{Dice Score (\%) ↑}}}                                                                                                                                   & \multicolumn{4}{c}{{\color[HTML]{000000} \textbf{Hausdorff Dist. (mm) ↓}}}                                                                                                                                \\ \cline{2-9} 
    \multirow{-2}{*}{{\color[HTML]{000000} \textbf{Method}}} & {\color[HTML]{000000} \hspace{-0.5em}\textbf{ET}}    & {\color[HTML]{000000} \hspace{-0.5em}\textbf{WT}}    & \multicolumn{1}{c|}{{\color[HTML]{000000} \hspace{-0.5em}\textbf{TC}}}    & \multicolumn{1}{c|}{{\color[HTML]{000000} \textbf{Avg.}}}  & {\color[HTML]{000000} \hspace{-0.5em}\textbf{ET}}     & {\color[HTML]{000000} \hspace{-0.5em}\textbf{WT}}     & \multicolumn{1}{c|}{{\color[HTML]{000000} \hspace{-0.5em}\textbf{TC}}}     & \multicolumn{1}{c}{{\color[HTML]{000000} \textbf{Avg.}}} \\ \hline
    {\color[HTML]{000000} UNet 3D \cite{3DUNET}}                          & {\color[HTML]{000000} 76.42}          & {\color[HTML]{000000} 84.10}          & \multicolumn{1}{c|}{{\color[HTML]{000000} 75.64}}          & \multicolumn{1}{c|}{{\color[HTML]{000000} 75.39}}          & {\color[HTML]{000000} 68.722}          & {\color[HTML]{000000} 71.324}          & \multicolumn{1}{c|}{{\color[HTML]{000000} 54.225}}          & {\color[HTML]{000000} 64.757}                             \\
    {\color[HTML]{000000} Attention UNet \cite{2018attentionunet}}                   & {\color[HTML]{000000} 79.11}          & {\color[HTML]{000000} 80.58}          & \multicolumn{1}{c|}{{\color[HTML]{000000} 80.22}}          & \multicolumn{1}{c|}{{\color[HTML]{000000} 79.97}}          & {\color[HTML]{000000} 49.628}          & {\color[HTML]{000000} 47.021}          & \multicolumn{1}{c|}{{\color[HTML]{000000} 47.388}}          & {\color[HTML]{000000} 48.012}                             \\
    {\color[HTML]{000000} Res-UNet 3D \cite{zhang2018road}}                   & {\color[HTML]{000000} 83.25}          & {\color[HTML]{000000} 82.39}          & \multicolumn{1}{c|}{{\color[HTML]{000000} 81.32}}          & \multicolumn{1}{c|}{{\color[HTML]{000000} 82.32}}          & {\color[HTML]{000000} 46.648}          & {\color[HTML]{000000} 47.288}          & \multicolumn{1}{c|}{{\color[HTML]{000000} 45.892}}          & {\color[HTML]{000000} 46.609} \\
    {\color[HTML]{000000} ConResNet\cite{zhang2020inter}}                            & {\color[HTML]{000000} 83.58}          & {\color[HTML]{000000} 83.62}          & \multicolumn{1}{c|}{{\color[HTML]{000000} 84.19}}          & \multicolumn{1}{c|}{{\color[HTML]{000000} 83.79}}          & {\color[HTML]{000000} 43.208}          & {\color[HTML]{000000} 45.572}          & \multicolumn{1}{c|}{{\color[HTML]{000000} 47.286}}          & {\color[HTML]{000000} 45.355}                             \\
    {\color[HTML]{000000} nnUNet \cite{2021nnunet}}                            & {\color[HTML]{000000} 83.33}          & {\color[HTML]{000000} 83.70}          & \multicolumn{1}{c|}{{\color[HTML]{000000} 80.24}}          & \multicolumn{1}{c|}{{\color[HTML]{000000} 84.42}}          & {\color[HTML]{000000} 45.922}          & {\color[HTML]{000000} 43.680}          & \multicolumn{1}{c|}{{\color[HTML]{000000} 52.177}}          & {\color[HTML]{000000} 47.260}                             \\
    {\color[HTML]{000000} TransUNet \cite{chen2021transunet}}                         & {\color[HTML]{000000} 82.57}          & {\color[HTML]{000000} 82.01}          & \multicolumn{1}{c|}{{\color[HTML]{000000} 83.12}}          & \multicolumn{1}{c|}{{\color[HTML]{000000} 82.57}}          & {\color[HTML]{000000} 49.212}          & {\color[HTML]{000000} 45.767}          & \multicolumn{1}{c|}{{\color[HTML]{000000} 42.159}}          & {\color[HTML]{000000} 45.713}                             \\
    {\color[HTML]{000000} TransBTS \cite{wenxuan2021transbts}}                          & {\color[HTML]{000000} 85.44}          & {\color[HTML]{000000} 84.61}          & \multicolumn{1}{c|}{{\color[HTML]{000000} 82.90}}          & \multicolumn{1}{c|}{{\color[HTML]{000000} 84.32}}          & {\color[HTML]{000000} 38.051}          & {\color[HTML]{000000} 38.872}          & \multicolumn{1}{c|}{{\color[HTML]{000000} 42.112}}          & {\color[HTML]{000000} 39.678}                             \\
    {\color[HTML]{000000} VT-UNet \cite{peiris2022robust}}                           & {\color[HTML]{000000} 86.10}          & {\color[HTML]{000000} 87.22}          & \multicolumn{1}{c|}{{\color[HTML]{000000} 84.33}}          & \multicolumn{1}{c|}{{\color[HTML]{000000} 85.88}}          & {\color[HTML]{000000} 35.794}          & {\color[HTML]{000000} 34.321}          & \multicolumn{1}{c|}{{\color[HTML]{000000} 38.210}}          & {\color[HTML]{000000} 36.078}                             \\
    {\color[HTML]{000000} PANet \cite{PANet}}                             & {\color[HTML]{000000} 85.43}          & {\color[HTML]{000000} 86.04}          & \multicolumn{1}{c|}{{\color[HTML]{000000} 84.71}}          & \multicolumn{1}{c|}{{\color[HTML]{000000} 85.39}}          & {\color[HTML]{000000} 39.720}          & {\color[HTML]{000000} 37.155}          & \multicolumn{1}{c|}{{\color[HTML]{000000} 42.010}}          & {\color[HTML]{000000} 39.628}                             \\
    {\color[HTML]{000000} UNETR++ \cite{shaker2024unetr++}}                           & {\color[HTML]{000000} 87.76}          & {\color[HTML]{000000} 89.68}          & \multicolumn{1}{c|}{{\color[HTML]{000000} 85.42}}          & \multicolumn{1}{c|}{{\color[HTML]{000000} 87.62}}          & {\color[HTML]{000000} 33.818}          & {\color[HTML]{000000} 20.341}          &\multicolumn{1}{c|} {\color[HTML]{000000} 21.821}                               & {\color[HTML]{000000} 25.327} \\
    {\color[HTML]{000000} UNETR \cite{2022unetr}}                             & {\color[HTML]{000000} 87.37}          & {\color[HTML]{000000} 89.53}          & \multicolumn{1}{c|}{{\color[HTML]{000000} 86.29}}          & \multicolumn{1}{c|}{{\color[HTML]{000000} 87.73}}          & {\color[HTML]{000000} 31.237}          & {\color[HTML]{000000} 28.638}          & \multicolumn{1}{c|}{{\color[HTML]{000000} 32.109}}          & {\color[HTML]{000000} 30.661}                             \\
    {\color[HTML]{000000} S$^2$CA-Net \cite{zhou2024shape}}                          & {\color[HTML]{00009B} \textbf{88.87}} & {\color[HTML]{00009B} \textbf{91.55}} & \multicolumn{1}{c|}{{\color[HTML]{00009B} \textbf{87.00}}} & \multicolumn{1}{c|}{{\color[HTML]{00009B} \textbf{89.14}}} & {\color[HTML]{00009B} \textbf{32.216}} & {\color[HTML]{00009B} \textbf{10.813}} & \multicolumn{1}{c|}{{\color[HTML]{00009B} \textbf{17.035}}} & {\color[HTML]{00009B} \textbf{20.021}}                    \\ \cline{2-9}  \hline
    {\color[HTML]{000000} \textbf{HFF-Net(Ours)}}            & {\color[HTML]{FE0000} \textbf{96.16}}      & {\color[HTML]{FE0000} \textbf{95.56}} & \multicolumn{1}{c|}{{\color[HTML]{FE0000} \textbf{96.34}}}      & \multicolumn{1}{c|}{{\color[HTML]{FE0000} \textbf{96.02}}}      & {\color[HTML]{FE0000} \textbf{5.576}}       & {\color[HTML]{FE0000} \textbf{6.272}}  & \multicolumn{1}{c|}{{\color[HTML]{FE0000} \textbf{6.432}}}       & {\color[HTML]{FE0000} \textbf{6.093}}                          \\ \hline
\end{tabular}}
    \end{subtable}
    \vspace{-8pt}
    \begin{subtable}
        \centering
        \caption{Comparison of Segmentation Performance on MSD BTS Dataset. The Best Two Results Are Shown In \textcolor[HTML]{FE0000}{Red} And \textcolor[HTML]{00009B}{Blue} Fonts, Respectively}
        \label{tab:tab2}
        \resizebox*{1\linewidth}{0.16\textheight}{
    \begin{tabular}{c|cccc|cccc}
    \hline
     & \multicolumn{4}{c|}{\textbf{Dice Score (\%) ↑}} & \multicolumn{4}{c}{ \textbf{Hausdorff Dist. (mm) ↓}} \\ \cline{2-9} 
    \multirow{-2}{*}{\textbf{Method}} & \textbf{ET} & \textbf{WT} & \multicolumn{1}{c|}{\textbf{TC}} & \textbf{Avg.} & \textbf{ET} & \textbf{WT} & \multicolumn{1}{c|}{\textbf{TC}} & \multicolumn{1}{c}{\textbf{Avg.}} \\ \hline
    UNet 3D \cite{3DUNET} & 56.14 & 76.62 & \multicolumn{1}{c|}{66.51} & 66.42 & 11.118 & 9.212 & \multicolumn{1}{c|}{10.237} & 10.189 \\
    Attention UNet \cite{2018attentionunet} & 54.31 & 76.74 & \multicolumn{1}{c|}{68.32} & 66.46 & 10.451 & 9.121 & \multicolumn{1}{c|}{10.463} & 10.012 \\
    SETR NUP \cite{zheng2021rethinking} & 54.42 & 69.71 & \multicolumn{1}{c|}{66.91} & 63.68 & 11.721 & 14.419 & \multicolumn{1}{c|}{15.190} & 13.777 \\
    SETR PUP \cite{zheng2021rethinking} & 54.93 & 69.60 & \multicolumn{1}{c|}{67.00} & 63.84 & 11.762 & 15.245 & \multicolumn{1}{c|}{15.023} & 14.010 \\
    SETR MLA \cite{zheng2021rethinking} & 55.44 & 69.78 & \multicolumn{1}{c|}{66.51} & 63.91 & 10.236 & 15.503 & \multicolumn{1}{c|}{14.722} & 13.487 \\
    TransUNet \cite{chen2021transunet} & 54.19 & 70.62 & \multicolumn{1}{c|}{68.40} & 64.40 & 10.416 & 14.029 & \multicolumn{1}{c|}{14.501} & 12.982 \\
    TransBTS \cite{wenxuan2021transbts} & 57.43 & 77.91 & \multicolumn{1}{c|}{73.52} & 69.62 & 9.973 & 10.032 & \multicolumn{1}{c|}{8.951} & 9.652 \\
    CoTr \cite{xie2021cotr} & 55.74 & 74.57 & \multicolumn{1}{c|}{74.76} &  68.37 & {9.452} & {9.199} & \multicolumn{1}{c|}{10.448} & {9.700} \\
    UNETR \cite{2022unetr} & 58.52 & 78.94 & \multicolumn{1}{c|} {76.11} & 71.19 & 9.349 & 8.271 & \multicolumn{1}{c|}{8.853} & 8.824 \\
    nnUNet \cite{2021nnunet} & 80.97 & 91.90 & \multicolumn{1}{c|}{85.35} & 86.07 & 4.058 & 3.637 & \multicolumn{1}{c|}{4.912} & 4.202 \\
    nnformer \cite{2021nnformer} & 81.82 & 91.31 & \multicolumn{1}{c|}{85.97} & 86.39 & 3.870 & 3.804 & \multicolumn{1}{c|}{4.492} & 4.055 \\
    VT-UNet \cite{peiris2022robust} & 82.20 & 91.90 & \multicolumn{1}{c|}{87.20} & 87.10 & 2.683 & 3.511 & \multicolumn{1}{c|}{4.092} & 3.429 \\
    S$^2$CA-Net \cite{zhou2024shape} & {\color[HTML]{00009B} \textbf{82.37}} & {\color[HTML]{00009B} \textbf{92.45}} & \multicolumn{1}{c|} {\color[HTML]{00009B} \textbf{88.91}} & {\color[HTML]{00009B} \textbf{87.91}} & 2.712 & 3.043 & \multicolumn{1}{c|}{3.877} & 3.211 \\ \hline
    \textbf{HFF-Net(Ours)} & {\color[HTML]{FE0000} \textbf{87.28}} & {\color[HTML]{FE0000} \textbf{94.03}} & \multicolumn{1}{c|}{{\color[HTML]{FE0000} \textbf{89.09}}} & {\color[HTML]{FE0000} \textbf{90.01}} & 2.885 & 4.790 & \multicolumn{1}{c|}{6.067} & 4.580 \\ \hline
     \end{tabular}}
    \end{subtable}
    \vspace{-8pt}
    \begin{subtable}
        \centering
        \caption{Comparison of Segmentation Performance, Model Size, and Computational Complexity on BRATS2020 Dataset. Best Two Results Are Shown in \textcolor[HTML]{FE0000}{Red} and \textcolor[HTML]{00009B}{Blue}, Respectively}
        \label{tab:tab3}
        \resizebox*{0.95\linewidth}{0.185\textheight}{
        \begin{tabular}{c|cccc|clcl}
        \hline
                                      & \multicolumn{4}{c|}{\textbf{Dice Score (\%) ↑}}                                                                                                                                    & \multicolumn{2}{c|}{{\color[HTML]{000000} }}                                        & \multicolumn{2}{c}{}                                    \\ \cline{2-5}
    \multirow{-2}{*}{\textbf{Method}} & \textbf{ET}                           & \textbf{WT}                           & \multicolumn{1}{c|}{\textbf{TC}}                           & \textbf{Avg.}                         & \multicolumn{2}{c|}{\multirow{-2}{*}{{\color[HTML]{000000} \textbf{Param. (M) ↓}}}} & \multicolumn{2}{c}{\multirow{-2}{*}{\textbf{GFlops ↓}}} \\ \hline
    UNet 3D \cite{3DUNET}                          & 68.76                                 & 84.11                                 & \multicolumn{1}{c|}{79.06}                                 & 77.31                                 & \multicolumn{2}{c}{16.09}                                                           & \multicolumn{2}{c}{586.71}                              \\
    V-Net \cite{diceloss}                            & 68.97                                 & 86.11                                 & \multicolumn{1}{c|}{77.90}                                 & 77.66                                 & \multicolumn{2}{c}{69.30}                                                           & \multicolumn{2}{c}{765.90}                              \\
    Attention UNet \cite{2018attentionunet}                   & 75.90                                 & 90.50                                 & \multicolumn{1}{c|}{79.60}                                 & 82.00                                 & \multicolumn{2}{c}{17.12}                                                           & \multicolumn{2}{c}{588.26}                              \\
    Residual UNet 3D \cite{yu2019liver}                & 71.63                                 & 82.46                                 & \multicolumn{1}{c|}{76.47}                                 & 76.85                                 & \multicolumn{2}{c}{--}                                                              & \multicolumn{2}{c}{--}                                  \\
    SGEResU-Net \cite{liu2022sgeresu}                       & 77.48                                 & 89.91                                 & \multicolumn{1}{c|}{81.04}                                 & 82.81                                 & \multicolumn{2}{c}{{\color[HTML]{000000} --}}                                       & \multicolumn{2}{c}{{\color[HTML]{000000} --}}           \\
    Cascaded U-Net \cite{jiang2020two}                   & 73.60                                 & 90.80                                 & \multicolumn{1}{c|}{81.00}                                 & 81.80                                 & \multicolumn{2}{c}{33.8}                                                         & \multicolumn{2}{c}{1173.22}                               \\
    nnUNet \cite{2021nnunet}                             & 78.89                                 & 91.24                                 & \multicolumn{1}{c|}{85.06}                                 & 85.40                                 & \multicolumn{2}{c}{28.50}                                                           & \multicolumn{2}{c}{{\color[HTML]{000000} 1449.59}}      \\
    H$^2$NF-Net \cite{jia2021hnf}                          & 78.49                                 & 91.26                                 & \multicolumn{1}{c|}{83.53}                                 & 84.43                                 & \multicolumn{2}{c}{26.07}                                                           & \multicolumn{2}{c}{621.09}                              \\
    TransUNet \cite{chen2021transunet}                         & 78.42                                 & 89.46                                 & \multicolumn{1}{c|}{78.37}                                 & 82.08                                 & \multicolumn{2}{c}{105.28}                                                          & \multicolumn{2}{c}{1205.76}                             \\
    Swin-UNet (2D) \cite{swin}                    & 74.00                                 & 87.20                                 & \multicolumn{1}{c|}{80.90}                                 & 80.83                                 & \multicolumn{2}{c}{27.17}                                                           & \multicolumn{2}{c}{357.49}                              \\
    TransBTS \cite{wenxuan2021transbts}                          & 78.73                                 & 90.09                                 & \multicolumn{1}{c|}{81.71}                                 & 83.52                                 & \multicolumn{2}{c}{32.99}                                                           & \multicolumn{2}{c}{333.00}                              \\
    UNETR \cite{2022unetr}                             & 78.80                                 & 89.90                                 & \multicolumn{1}{c|}{{\color[HTML]{000000} 84.20}}          & 84.30                                 & \multicolumn{2}{c}{92.58}                                                           & \multicolumn{2}{c}{41.19}                               \\
    UNeXt \cite{valanarasu2022unext}                             & 76.49                                 & 88.70                                 & \multicolumn{1}{c|}{81.37}                                 & 82.19                                 & \multicolumn{2}{c}{1.47}                                                            & \multicolumn{2}{c}{0.45}                                \\
    PANet \cite{PANet}                             & 78.40                                 & 90.90                                 & \multicolumn{1}{c|}{83.10}                                 & 84.13                                 & \multicolumn{2}{c}{19.23}                                                           & \multicolumn{2}{c}{609.98}                              \\
    TranBTSV2 \cite{li2022transbtsv2}                         & 79.63                                 & 90.56                                 & \multicolumn{1}{c|}{84.50}                                 & 84.90                                 & \multicolumn{2}{c}{15.30}                                                           & \multicolumn{2}{c}{241.00}                              \\
    S$^2$CA-Net \cite{zhou2024shape}                          & {\color[HTML]{00009B} \textbf{80.41}} & {\color[HTML]{00009B} \textbf{91.37}} & \multicolumn{1}{c|}{{\color[HTML]{00009B} \textbf{85.21}}} & {\color[HTML]{00009B} \textbf{85.66}} & \multicolumn{2}{c}{21.41}                                                           & \multicolumn{2}{c}{113.47}                              \\ \cline{2-9}   \hline
    \textbf{HFF-Net(Ours)}            & {\color[HTML]{FE0000} \textbf{87.36}} & {\color[HTML]{FE0000} \textbf{92.37}} & \multicolumn{1}{c|}{{\color[HTML]{FE0000} \textbf{88.22}}}                      & {\color[HTML]{FE0000} \textbf{89.31}} & \multicolumn{2}{c}{36.01}                                                           & \multicolumn{2}{c}{541.18}                              \\ \hline
    \end{tabular}}
    \end{subtable}
    \vspace{12pt}
    \begin{subtable}
    \centering
    \caption{Comparison of Segmentation Performance on BRATS2019 Dataset. \textcolor[HTML]{FE0000}{Red} Font Denotes the Best Results And \textcolor[HTML]{00009B}{Blue} Font Represents The Second Best Results}
\label{tab:tab4}
\resizebox*{\linewidth}{0.172\textheight}{
\begin{tabular}{c|cccccccc}
\hline
{\color[HTML]{000000} }                                  & \multicolumn{4}{c|}{{\color[HTML]{000000} \textbf{Dice Score (\%) $\uparrow$}}}                                                                                                                                   & \multicolumn{4}{c}{{\color[HTML]{000000} \textbf{Hausdorff Dist. (mm) $\downarrow$}}}                                                                                    \\ \cline{2-9} 
\multirow{-2}{*}{\textbf{Method}} & \textbf{ET}                           & \textbf{WT}                           & \multicolumn{1}{c|}{\textbf{TC}}                           & \multicolumn{1}{c|}{\textbf{Avg.}}                         & \textbf{ET}                           & \textbf{WT}                           & \multicolumn{1}{c|}{\textbf{TC}}                           & \multicolumn{1}{c}{\textbf{Avg.}} \\ \hline
UNet 3D \cite{3DUNET}                          & 70.86                                 & 87.38                                 & \multicolumn{1}{c|}{72.48}                                 & \multicolumn{1}{c|}{76.91}                                 & 5.062                                 & 9.432                                 & \multicolumn{1}{c|}{8.719}                                 & 7.738                              \\
V-Net \cite{diceloss}                             & 73.89                                 & 88.73                                 & \multicolumn{1}{c|}{76.56}                                 & \multicolumn{1}{c|}{79.73}                                 & 6.131                                 & 6.256                                 & \multicolumn{1}{c|}{8.705}                                 & 7.031                              \\
Attention UNet \cite{2018attentionunet}                   & 75.96                                 & 88.81                                 & \multicolumn{1}{c|}{77.20}                                 & \multicolumn{1}{c|}{80.66}                                 & 5.202                                 & 7.756                                 & \multicolumn{1}{c|}{8.258}                                 & 7.702                              \\
Bag of tricks \cite{zhao2020bag}                     & 70.20                                 & 89.30                                 & \multicolumn{1}{c|}{80.00}                                 & \multicolumn{1}{c|}{79.83}                                 & 4.766                                 & 5.078                                 & \multicolumn{1}{c|}{6.472}                                 & 5.439                              \\
Myronenko et al. \cite{myronenko2020robust}                  & 80.00                                 & 89.40                                 & \multicolumn{1}{c|}{83.40}                                 & \multicolumn{1}{c|}{84.27}                                 &  3.921          &  5.890        & \multicolumn{1}{c|}{ 6.562}          & 5.458                              \\
Tunet \cite{vu2020tunet}                             & 78.42                                 & 90.34                                 & \multicolumn{1}{c|}{81.12}                                 & \multicolumn{1}{c|}{83.29}                                 & 3.700                                 & 4.320                                 & \multicolumn{1}{c|}{6.280}                                 & 4.767                              \\
3D KiU-Net \cite{valanarasu2020kiu}                        & 73.21                                 & 87.60                                 & \multicolumn{1}{c|}{73.92}                                 & \multicolumn{1}{c|}{78.24}                                 & 6.323                                 & 8.942                                 & \multicolumn{1}{c|}{ 9.893}          & 8.386                              \\
TransUNet \cite{chen2021transunet}                         & 78.17                                 & 89.48                                 & \multicolumn{1}{c|}{78.91}                                 & \multicolumn{1}{c|}{82.19}                                 & 4.832                                 & 6.667                                 & \multicolumn{1}{c|}{7.365}                                 & 6.288                              \\
Swin-UNet (2D) \cite{swin}                    & 78.49                                 & 89.38                                 & \multicolumn{1}{c|}{78.75}                                 & \multicolumn{1}{c|}{82.21}                                 & 7.505                                 & 6.925                                 & \multicolumn{1}{c|}{9.260}                                 & 7.897                              \\
TransBTS \cite{wenxuan2021transbts}                         & 78.93                                 & 90.00                                 & \multicolumn{1}{c|}{81.94}                                 & \multicolumn{1}{c|}{83.62}                                 & 3.736                                 & 5.644                                 & \multicolumn{1}{c|}{6.049}                                 & 5.143                              \\
SA-LuT-Nets \cite{yu2021sa}                       & 78.21                                 & 90.79                                 & \multicolumn{1}{c|}{{\color[HTML]{00009B} \textbf{84.82}}} & \multicolumn{1}{c|}{84.61}                                 & 3.690                                 & 4.460                                 & \multicolumn{1}{c|}{{\color[HTML]{FE0000} \textbf{5.260}}} & 4.470                              \\
Med-DANet \cite{wang2022med}                         & 79.99                                 & 90.13                                 & \multicolumn{1}{c|}{80.83}                                 & \multicolumn{1}{c|}{83.65}                                 & 4.086                                 & 5.826                                 & \multicolumn{1}{c|}{6.886}                                 & 5.599                              \\
PANet \cite{PANet}                             & 78.17                                 & 90.54                                 & \multicolumn{1}{c|}{82.98}                                 & \multicolumn{1}{c|}{83.90}                                 & {\color[HTML]{00009B} \textbf{3.453}} & {\color[HTML]{00009B} \textbf{4.975}} & \multicolumn{1}{c|}{6.852}                                 & 5.093                              \\
S$^2$CA-Net \cite{zhou2024shape}                          & {\color[HTML]{00009B} \textbf{80.14}} & {\color[HTML]{00009B} \textbf{90.97}} & \multicolumn{1}{c|}{84.35}                                 & \multicolumn{1}{c|}{85.15}                                 & {\color[HTML]{FE0000} \textbf{3.177}} & {\color[HTML]{FE0000} \textbf{4.001}} & \multicolumn{1}{c|}{{\color[HTML]{00009B} \textbf{5.860}}} & 4.346                              \\ \hline
\textbf{HFF-Net(Ours)}            & {\color[HTML]{FE0000} \textbf{85.35}} & {\color[HTML]{FE0000} \textbf{91.58}} & \multicolumn{1}{c|}{{\color[HTML]{FE0000} \textbf{87.59}}} & \multicolumn{1}{c|}{{\color[HTML]{FE0000} \textbf{88.17}}} & 5.067                                 & 5.301                                 & \multicolumn{1}{c|}{6.033}                                 & 5.467                              \\ \hline
\end{tabular}}
    \end{subtable}
\end{table}
\subsubsection{Quantitative Results} 
We thoroughly evaluated our proposed HFF-Net on four publicly available brain tumor segmentation datasets: BRATS2023-MEN, MSD BTS Task, BraTS2020, and BraTS2019, with all results summarized in Tables~\Cref{tab:tab1,tab:tab2,tab:tab3,tab:tab4}. To ensure a fair and robust assessment, we adopted five-fold cross-validation or subject-wise hold-out splitting across BraTS2019 and BraTS2020 datasets, with training and validation subjects strictly separated at the patient level to avoid data leakage and validation bias. We compared our results with the results referenced from S$^2$CA-Net \cite{zhou2024shape}, a recent state-of-the-art method that employs a patch-based inference strategy and post-processing to suppress false positives. In contrast, our model performs end-to-end inference without any post-processing. Its design, which combines full-volume processing with a dual-branch frequency-aware architecture, enables effective capture of brain tumor characteristics, contributing to its superior performance over other methods.
\begin{table*}[!t]
\centering
\caption{Additional significance tests on recent SOTA methods for segmentation performance on four datasets, reported as mean ± std. * indicate P-value < 0.05.}
\label{tab:tab6 sig}
\begin{adjustbox}{width=0.8\linewidth} 
\renewcommand{\arraystretch}{1.1}
\begin{tabular}{c|c|c|ccc|ccc}
\hline
\multirow{2}{*}{\textbf{Dataset}} 
 & \multirow{2}{*}{\textbf{Method}} 
 & \multirow{2}{*}{\textbf{Mark}} 
 & \multicolumn{3}{c|}{\hspace{1em}\textbf{Dice (\%)  {$\uparrow$}}} 
 & \multicolumn{3}{c}{\hspace{1em}\textbf{HD95 (mm)  {$\downarrow$}}} \\

\cline{4-9}
 &  &  
 & \textbf{ET} & \textbf{WT} & \textbf{TC}
 & \textbf{ET} & \textbf{WT} & \textbf{TC} \\
\hline

\multirow{4}{*}{\makecell{MSD\\-BTS}} 
 & \hspace{1em}UNETR \cite{2022unetr} & WACV22 
 &\hspace{0.5em}60.5±2.3*   &\hspace{0.5em}79.3±1.4*  &\hspace{0.5em}77.1±1.8*  & \hspace{0.5em}9.3±1.5*  &\hspace{0.5em}9.0±1.0*  &\hspace{0.5em}9.5±1.2*  \\
 & \hspace{1em}T-UNet \cite{peiris2022robust} & MICCAI22 
 & \hspace{0.5em}81.3±0.1* &\hspace{0.5em}91.5±0.2*  &87.4±0.4  &2.7±0.3  &4.5±1.2  & 4.3±0.6  \\
 & \hspace{1em}E2ENet \cite{wu2024e2enet} & NeurIPS24
 & \hspace{0.5em}71.5±2.6* & \hspace{0.5em}78.9±3.3* & \hspace{0.5em}73.1±4.9* 
 & \hspace{0.5em}11.5±0.3* & \hspace{0.5em}13.2±1.6* & \hspace{0.5em}14.5±2.1* \\
\rowcolor{gray!15}
\multicolumn{1}{c|}{\cellcolor{white}} 
 & HFF-Net & Ours
 &87.1±0.2  &94.2±0.3  & 89.4±0.3 
 &2.5±0.4  &4.1±0.6  &5.8±0.7  \\
\hline

\multirow{4}{*}{BraTS19}
  & \hspace{0.5em}SwinUNETR \cite{2021swinunetr} & arXiv21
 & \hspace{0.5em}79.7±3.8* & \hspace{0.5em}87.9±1.2* & \hspace{0.5em}82.5±0.9* 
 & \hspace{0.5em}11.7±0.1* & \hspace{0.5em}8.5±2.1* & 7.1±1.9 \\
  & \hspace{1em}PANet \cite{PANet} & TMI22
 &\hspace{0.5em}77.5±0.5*  &89.8±1.9  &\hspace{0.5em}82.4±0.1*  &3.5±0.5  &4.7±0.7  &6.3±4.3  \\
 & \hspace{0.5em}MedNeXt \cite{roy2023mednext} & MICCAI23
 & \hspace{0.5em}80.5±1.2* &90.9±0.9  &\hspace{0.5em}83.8±0.3* &6.8±0.1  &\hspace{0.5em}10.8±0.2*  &8.9±0.1  \\
\rowcolor{gray!15}
\multicolumn{1}{c|}{\cellcolor{white}}
 & HFF-Net & Ours
 & 85.3±0.4 & 91.6±0.3 & 87.9±0.2
 & 5.2±0.2 & 5.9±0.1 & 6.3±1.1 \\
\hline

\multirow{4}{*}{BraTS20}
 & \hspace{1em}PANet \cite{PANet} & TMI22
 &\hspace{0.5em}78.8±3.7*  & 90.4±0.8 & \hspace{0.5em}83.2±1.5* & \hspace{0.5em}17.8±5.1*  & \hspace{0.5em}10.4±2.1* & \hspace{0.5em}9.7±1.9* \\
 & \hspace{1em}UNeXt \cite{valanarasu2022unext} & MICCAI22
 & \hspace{0.5em}76.2±6.3* &\hspace{0.5em}87.8±3.1*  & \hspace{0.5em}81.2±3.7* &\hspace{0.5em}10.5±2.5*  & \hspace{0.5em}9.2±4.1*  & 7.3±2.0 \\
 & \hspace{0em}CKD-TransBTS\cite{lin2023ckd} & TMI23
 &\hspace{0.5em}78.5±3.3*  & 91.4±1.4 & 85.7±1.0 & \hspace{0.5em}9.3±1.1*  & 5.9±0.6  &6.1±0.7  \\
\rowcolor{gray!15}
\multicolumn{1}{c|}{\cellcolor{white}}
 & HFF-Net & Ours
 & 87.1±0.2 & 92.3±0.1 & 88.1±0.4
 & 6.2±0.1 & 6.9±0.1 & 5.3±0.2 \\
\hline

\multirow{4}{*}{\makecell{BraTS23\\-MEN}}

 & \hspace{0em}SwinUNETRV2\cite{he2023swinunetr} & MICCAI23
 & \hspace{0.5em}85.4±2.3* &\hspace{0.5em}91.2±0.6*  &\hspace{0.5em}88.7±0.4*  & 6.2±0.3  & 5.4±0.2 & 6.0±0.3  \\
  & \hspace{1em}UNETR++ \cite{shaker2024unetr++} & TMI24 
 &\hspace{0.5em}87.5±3.1*  &\hspace{0.5em}89.4±0.8*  &\hspace{0.5em}85.6±0.9* & \hspace{0.5em}18.6±3.3*  & \hspace{0.5em}10.5±1.5*  & \hspace{0.5em}14.3±2.1* \\
 & \hspace{1em}SegMamba \cite{xing2024segmamba} & MICCAI24
 & \hspace{0.5em}87.2±1.2* & 93.4±0.1 & \hspace{0.5em}91.5±0.7*
 & 5.3±0.1 & 4.2±0.5 & 5.7±0.2 \\
\rowcolor{gray!15}
\multicolumn{1}{c|}{\cellcolor{white}}
 & HFF-Net & Ours
 & 96.1±0.2 & 95.3±0.1 & 96.1±0.2
 & 5.2±0.1 & 5.9±0.5 & 6.3±0.4 \\
\hline

\end{tabular}
\end{adjustbox}
    \vspace{-7pt}
\end{table*}
As shown in Table~\Cref{tab:tab1}, HFF-Net achieved Dice scores of 96.16\%, 95.56\%, and 96.34\% for ET (Enhancing Tumor), WT (Whole Tumor), and TC (Tumor Core) on BRATS2023-MEN, respectively, with corresponding HD95 distances of 5.576 mm, 6.272 mm, and 6.432 mm. The 8.29\% improvement in ET Dice score compared to S$^2$CA-Net indicates more accurate delineation of the enhancing tumor core, which is clinically critical for radiotherapy targeting and surgical planning, as it directly impacts the definition of high-dose regions and resection margins. On the MSD BTS dataset (Table~\Cref{tab:tab2}), HFF-Net achieved Dice scores of 87.28\%, 94.03\%, and 89.09\%, outperforming S$^2$CA-Net by 4.91\% in ET. This enhancement is especially relevant for reducing uncertainty in lesion localization, aiding in longitudinal monitoring and treatment response assessment. Similarly, Table~\Cref{tab:tab3} shows that HFF-Net reached Dice scores of 87.36\%, 92.37\%, and 88.22\% on BraTS2020, with an 8.65\% improvement in ET over S$^2$CA-Net. Despite the higher model complexity (36.01M parameters and 541.18 GFLOPs) due to its dual-branch design, the model maintains strong segmentation accuracy while offering improved delineation of tumor boundaries, which is essential for preoperative risk evaluation. On BraTS2019 (Table~\Cref{tab:tab4}), HFF-Net achieved Dice scores of 85.35\%, 91.58\%, and 87.59\%, outperforming S$^2$CA-Net by 6.5\% in ET, with slightly inferior HD95 results likely due to the smaller sample size. These results demonstrate consistent performance across diverse datasets and highlight HFF-Net’s robustness in segmenting clinically relevant tumor subregions, potentially enhancing clinical decision-making in both high-grade and low-grade glioma cases.

\textbf{Statistical Significance Analysis.}
To complement the above quantitative evaluations and strengthen the rigor of performance comparisons, we conducted additional statistical significance tests on four benchmark datasets, comparing HFF-Net with recent state-of-the-art methods. All experiments were conducted using five-fold cross-validation, and results are reported as mean ± standard deviation. This strategy helps reduce the influence of validation bias and ensures that the reported improvements reflect consistent performance across different data splits. As shown in Table \cref{tab:tab6 sig}, HFF-Net consistently outperforms other methods in Dice and HD95 across all tumor subregions. In particular, the enhancing tumor (ET) Dice scores show statistically significant improvements (p-value $<$ 0.05) on all datasets. For instance, on the BraTS23-MEN dataset, HFF-Net achieved an ET Dice of 96.1\%, compared to 87.2\% by SegMamba. Similarly, on MSD-BTS, our method reached 87.1\% in ET Dice, surpassing T-UNet (81.3\%) and E2ENet (71.5\%).

Beyond this, HFF-Net also shows strong performance in WT and TC segmentation. For example, on BraTS20, HFF-Net achieved TC Dice of 88.1\% with HD95 of 5.3 mm, while CKD-TransBTS yielded 85.7\% Dice and 6.1 mm HD95. On the more challenging BraTS19 dataset, our method still obtained consistent improvements in all subregions over SwinUNETR and MedNeXt. These results confirm the statistical significance of HFF-Net’s superior performance across all datasets, especially in the enhancing tumor region, where it consistently outperforms recent methods by a clear margin under standard evaluation protocols.

\textbf{Cross-Dataset Evaluation.}  
To assess the generalization ability of HFF-Net beyond brain tumor segmentation, we further evaluated it on two widely used datasets from other anatomical regions and modalities. The LA dataset (2018 Atrial Segmentation Challenge) \cite{xiong2021global} contains 100 high-resolution 3D cardiac MRI volumes, we used 80 for training and 20 for testing following~\cite{luo2021semi}, with a patch size of 96×96×80 and $\lambda_{\text{max}}$ set to 5. The LiTS dataset (2017 Liver Tumor Segmentation Challenge) \cite{bilic2023liver} comprises 131 abdominal CT scans, split into 100 training and 31 testing cases as in~\cite{wang2022rethinking}; we applied a soft-tissue window of [-100, 250] HU, cropped around the liver, and used a patch size of 112×112×32 with $\lambda_{\text{max}}$ set to 0.5. These datasets were selected to assess the model’s robustness across different imaging modalities and anatomies, which are among the most clinically relevant segmentation tasks beyond the brain.

\begin{figure}[!ht]
        \includegraphics[width=\columnwidth]{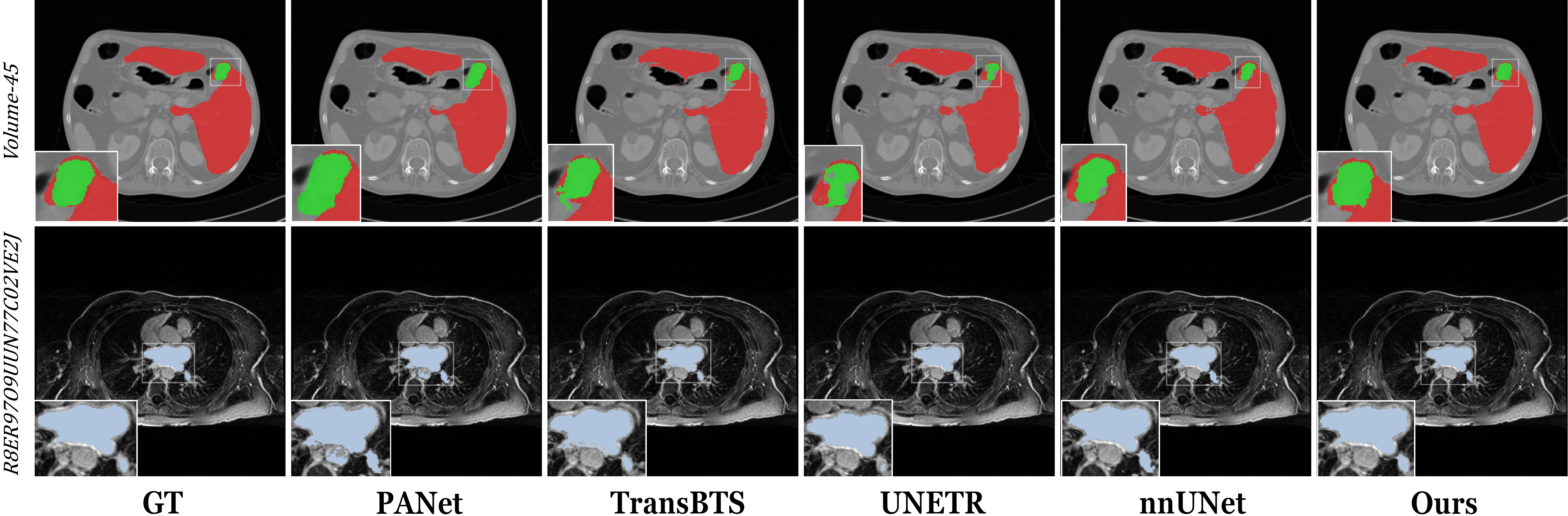}
    \vspace{-10pt}
    \caption{Segmentation results on the LiTS and LA datasets to show the generalizability of our method across different medical imaging modalities. In the LiTS dataset (upper row), \textcolor[HTML]{FE0000}{red} masks denote the liver and \textcolor[HTML]{00FF00}{green} masks represent the lesion region; in the LA dataset (lower row), \textcolor[HTML]{ADD8E6}{blue} masks indicate the left atrial cavity. Zoom in for details.}
    \label{la lits} 

\end{figure}

As shown in \Cref{la lits}, HFF-Net produces more accurate and spatially coherent segmentation results compared to other methods. On the LiTS dataset, it better captures organ contours and small lesion boundaries, while on the LA dataset, it preserves the continuity of the thin-walled atrial structure. These visual results highlight HFF-Net’s effectiveness in handling anatomical targets of varying size and morphology. Quantitative results are summarized in Table \cref{tab:two_dataset_comparison}, where HFF-Net consistently achieves the highest Dice scores and the lowest HD95 across both datasets. On LiTS, it attained 89.94\% Dice and 12.73 mm HD95, outperforming nnUNet by 3.63\% and 10.54 mm. On LA, it reached 94.12\% Dice and 2.35 mm HD95, with respective improvements of 2.17\% and 2.75 mm. These results demonstrate the model’s strong adaptability across different modalities and anatomical regions, reinforcing its potential for broader clinical applications.

\begin{table}[t]

    \centering

    \caption{Comparison of Segmentation Performance on LiTS and LA Datasets. 
    \textcolor[HTML]{FE0000}{Red} font denotes the best results and 
    \textcolor[HTML]{00009B}{Blue} font denotes the second best results.}
    \label{tab:two_dataset_comparison}
    \resizebox{1\linewidth}{!}{
    \begin{tabular}{c|cc|cc}
    \hline
    \multicolumn{1}{c|}{\multirow{2}{*}{\textbf{Method}}} 
    & \multicolumn{2}{c|}{\textbf{LiTS}} 
    & \multicolumn{2}{c}{\textbf{LA}} 
    \\ \cline{2-5}
    & \textbf{Dice (\%) ↑} & \textbf{HD95 (mm) ↓} 
    & \textbf{Dice (\%) ↑} & \textbf{HD95 (mm) ↓} 
    \\ \hline
    PANet \cite{PANet}
    & 77.17 & 45.16 
    & 85.42 & 10.85 
    \\ 
    TransBTS \cite{wenxuan2021transbts}
    & 84.52 & 24.91 
    & 89.33 & 9.23 
    \\ 
    UNETR \cite{2022unetr}
    & 78.29 & 36.02 
    & 88.71 & 8.76 
    \\ 
    nnUNet \cite{2021nnunet} 
    & \textcolor[HTML]{00009B}{\textbf{86.31}} & \textcolor[HTML]{00009B}{\textbf{23.27}}
    & \textcolor[HTML]{00009B}{\textbf{91.95}} & \textcolor[HTML]{00009B}{\textbf{5.10}}
    \\\hline
    \textbf{HFF-Net (Ours)}
    & \textcolor[HTML]{FE0000}{\textbf{89.94}} & \textcolor[HTML]{FE0000}{\textbf{12.73}}
    & \textcolor[HTML]{FE0000}{\textbf{94.12}} & \textcolor[HTML]{FE0000}{\textbf{2.35}}
    \\ \hline
    \end{tabular}
    }
\vspace{-3pt}
\end{table} 

\begin{figure}[!t]

    \vspace{4pt}
    \includegraphics[width=\columnwidth]{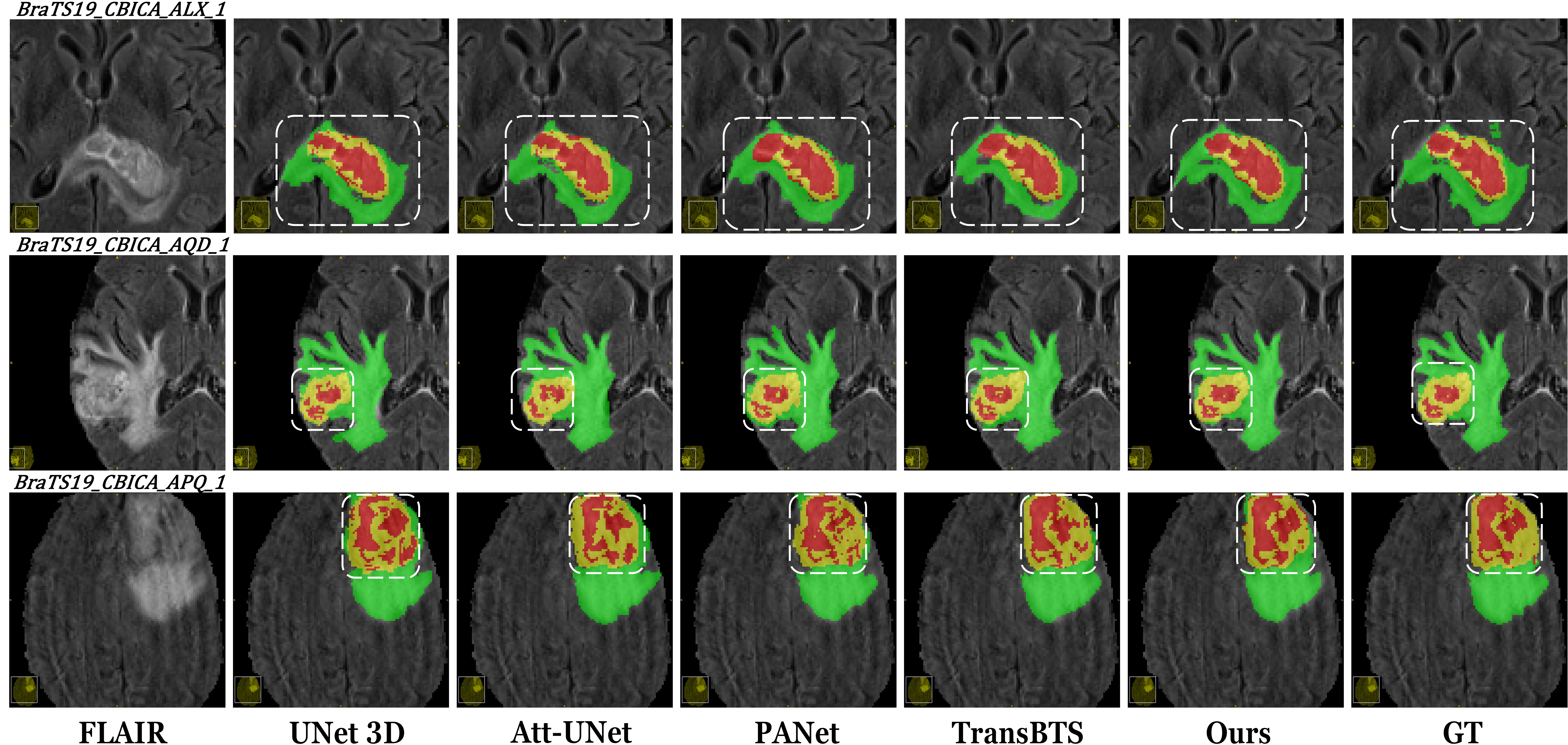}
     \vspace{-16pt}

    \caption{Segmentation performance comparison on the BraTS2019 training dataset. All cases feature structurally complex brain tumors with pronounced spatial heterogeneity. \textcolor[HTML]{FE0000}{Red} masks denote necrotic tumor core (NCR/NET), \textcolor[HTML]{FFD700}{yellow} masks denote enhancing tumors (ET), and \textcolor[HTML]{008000}{green} masks denote peritumoral edema (ED). White dotted boxes highlight brain tumor regions with significant segmentation differences.  Our method achieves more precise delineation, especially in cases with extensive ET involvement. Zoom in for details.}
    \label{Fig:2dc}

    \includegraphics[width=\columnwidth]{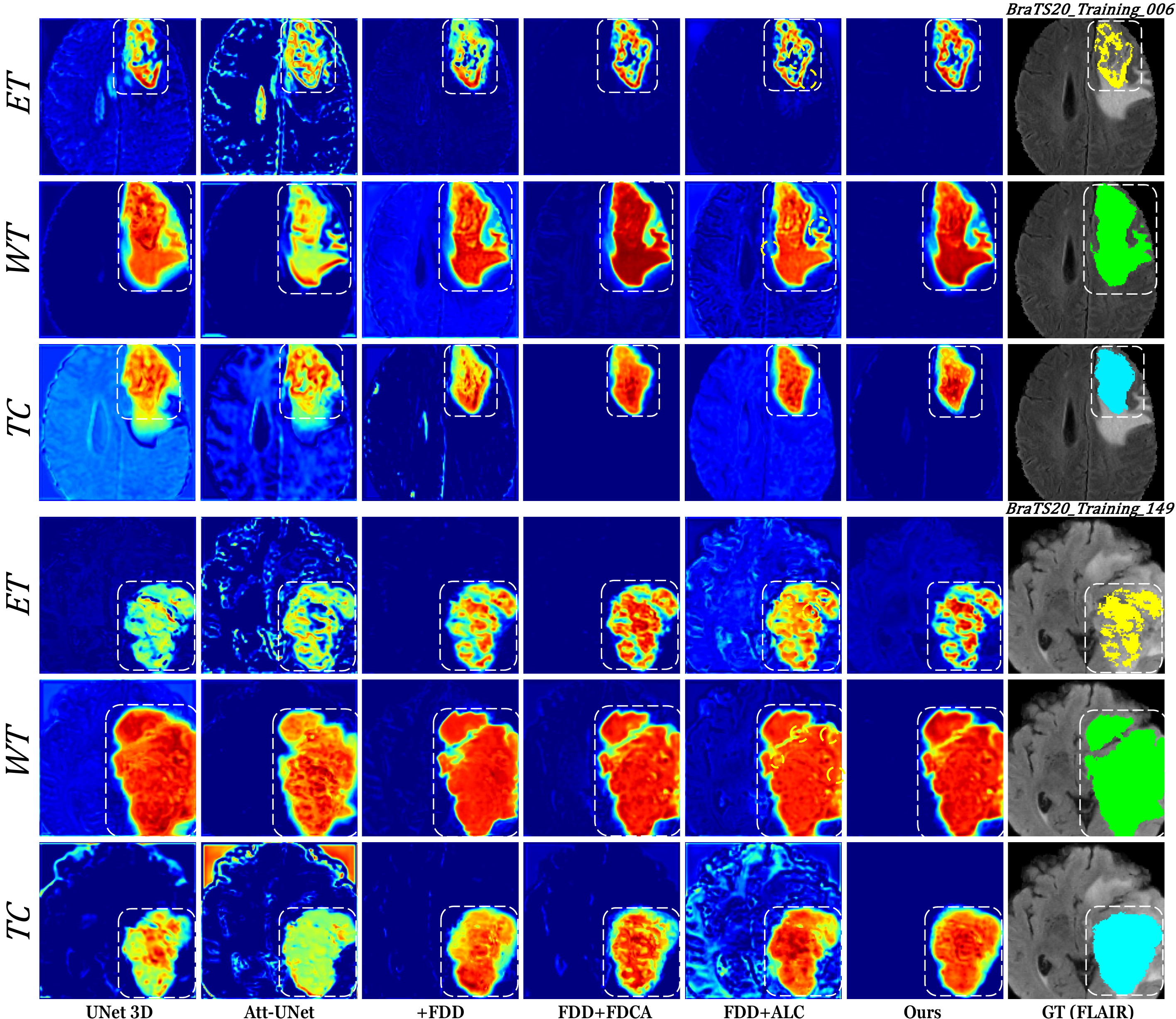} 
     \vspace{-16pt}

    \caption{ Grad-CAM visualization of different models on BraTS2020 training samples for ET, WT, and TC. \textcolor[HTML]{FE0000}{Red} indicates high attention, \textcolor[HTML]{00009B}{blue} low attention. Compared to baselines, HFF-Net variants show progressively sharper and more focused responses. The full model achieves the best boundary localization and fine-grained structural awareness, with \textcolor[HTML]{FFD700}{yellow} circles marking regions aligning closely with ground truth. Zoom in for details.}
    \label{Fig:att}

      \includegraphics[width=\columnwidth]{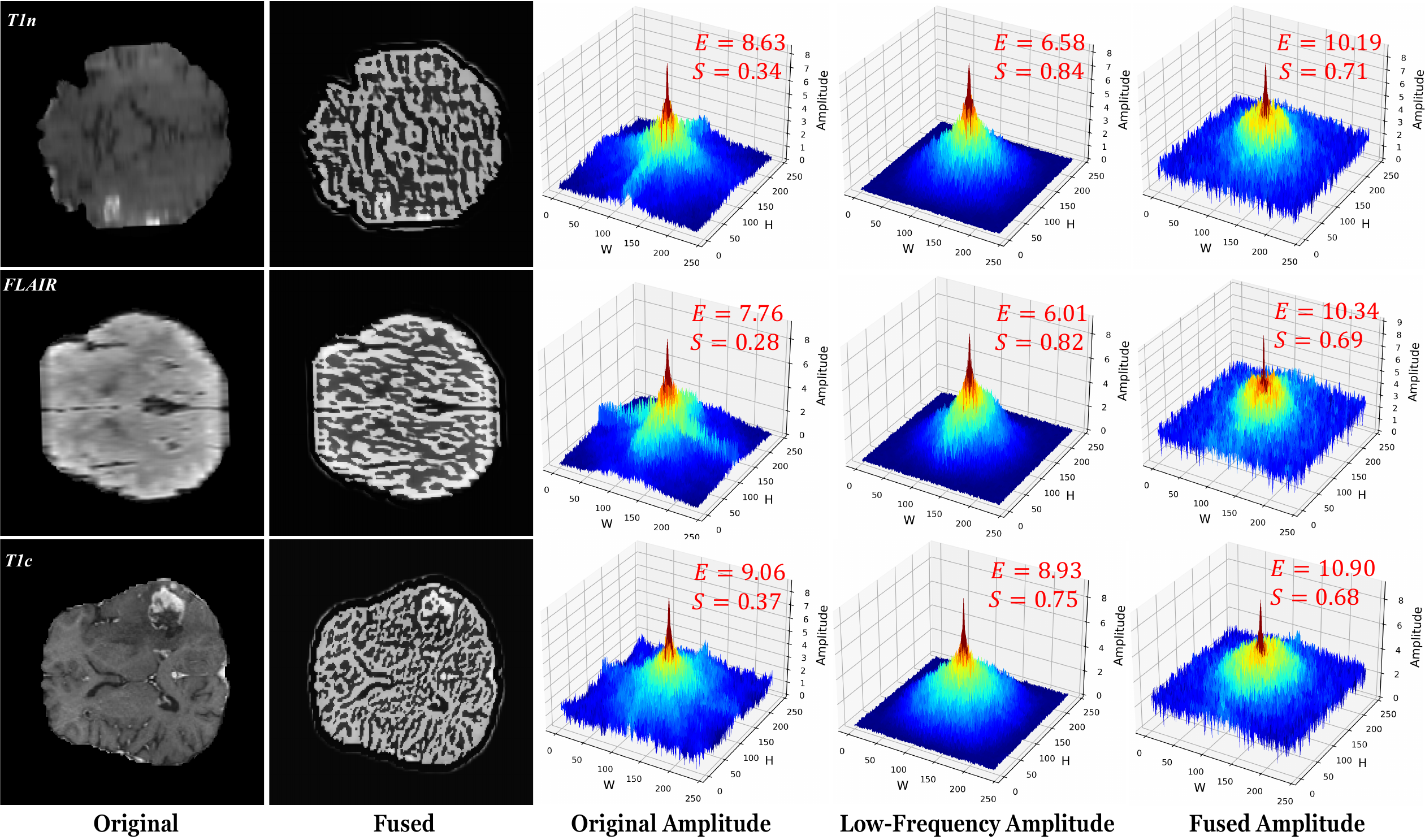} 
     \vspace{-16pt}

    \caption{Visualization of the frequency representations before and after DTCWT-based low-frequency decomposition and NSCT-based high-frequency fusion. The 3D amplitude plots illustrate how the frequency components are affected by these decomposition and fusion steps. The frequency-domain entropy (\( E\)) and shift-invariance (\( S\)) values, shown above each spectrum, quantify the representation of texture details and the robustness to image shifts. Zoom in for details.} 
    \label{Fig:amp} 
\vspace{-25pt}
\end{figure}

\begin{figure*}[t]
  \centering   \includegraphics[width=\linewidth]{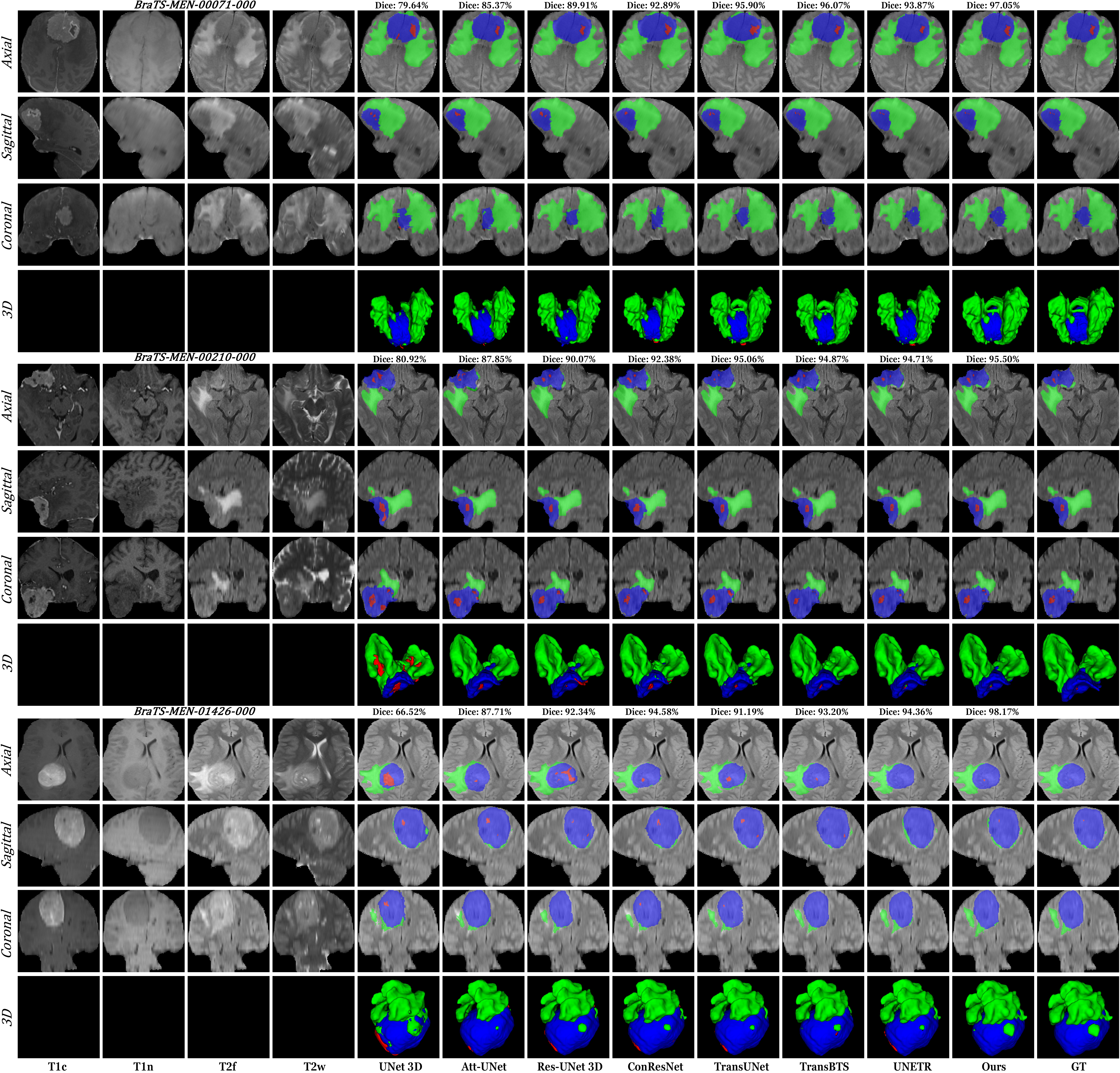}
   \vspace{-10pt}
   \caption{Visual comparison of segmentation results on axial, sagittal, coronal, and 3D views from the BraTS2023-MEN dataset. \textcolor[HTML]{FE0000}{Red} masks denote non-enhancing tumor core (NET), \textcolor[HTML]{00FF00}{green} masks represent surrounding non-enhancing FLAIR hyperintensity (SNFH), and \textcolor[HTML]{0000FF}{blue} masks denote enhancing tumor (ET). Dice scores above each column indicate the overall segmentation accuracy for each case. As this dataset generally features large ET regions, our method produces more anatomically faithful and spatially coherent 3D predictions compared to previous approaches. Zoom in for details.}
   \label{fig:fig7}  
   \vspace{-3pt}
\end{figure*}

\subsubsection{Qualitative Results} We evaluated the performance of HFF-Net through five-fold cross-validation on the BraTS2019 training set, comparing it with previous state-of-the-art methods. The qualitative results, depicted in \Cref{Fig:2dc}, showcase three samples with tumors at different locations. Each subsequent row represents an increasing level of segmentation difficulty. The images displayed in the axial plane include a global thumbnail at the bottom left, with white dashed boxes highlighting areas of pronounced segmentation differences. Consistent with the quantitative results, HFF-Net exhibits excellent performance in outlining overall tumor contours, particularly in handling highly complex ET regions due to the HF feature branch providing detailed edges and rich texture information. Notably, in the third row, where the tumor is the largest and the segmentation is most complex, our method significantly outperforms others in delineating the necrotic core and enhanced regions, demonstrating its robustness in intricate segmentation tasks. \Cref{fig:fig7} further compares results on the BraTS2023-MEN dataset, which predominantly contains cases with large enhancing tumor (ET) volumes. Dice scores above each column indicate overall segmentation accuracy per sample. Our model yields more anatomically complete and spatially coherent 3D predictions than prior methods, highlighting its robustness in challenging clinical scenarios.

To gain deeper insight into how our method achieves superior segmentation performance in complex tumor regions, we conducted a comparative analysis of Grad-CAM \cite{selvaraju2017grad} attention maps from three representative BraTS2020 samples, focusing on enhancing tumor (ET) areas. As shown in \Cref{Fig:att}, we visualized the attention distributions of UNet 3D, Att-UNet, and our HFF-Net under different module configurations. The first two networks exhibit dispersed attention with low-intensity responses in ET regions and blurry boundaries, indicating weak localization of key structures. The third column shows our model with only the FDD module, which yields stronger and more concentrated responses around tumor contours. However, the attention tends to over-accumulate within the lesion, leading to homogeneous activation that blurs boundary transitions in heterogeneous regions. Adding the FDCA module (4th column) intensifies attention across a broader spatial extent and enhances global contour coverage, but boundary sharpness remains suboptimal, and fine low-contrast details are still partially suppressed. In the fifth column, incorporating ALC results in slightly reduced overall intensity but improves spatial selectivity, enabling finer separation of small intratumoral lesions and thin or elongated structures. Notably, ALC occasionally highlights non-tumor areas, likely due to its increased sensitivity to high-frequency cues. While this facilitates the detection of fine structural details, it may also result in false positive activations in anatomically noisy regions, suggesting the need for additional regularization or uncertainty modeling in future work. The full model integrating FDD, FDCA, and ALC (6th column) exhibits balanced attention with sharp, well-localized focus on both global morphology and subtle structural variations, aligning closely with ground truth annotations and demonstrating the synergistic effect of all components.

\textbf{Frequency-Domain Analysis of Fusion Effectiveness.} To validate why combining DTCWT-based low-frequency (LF) decomposition and NSCT-based high-frequency (HF) fusion enhances segmentation performance, we conducted a frequency-domain analysis on three MRI samples from different modalities. The T1n and FLAIR images exhibit motion blur, while T1c serves as a clear reference. As shown in \Cref{Fig:amp}, each row displays, from left to right: the original image, LF and HF fused result, original amplitude spectrum, LF amplitude, and fused amplitude. We define two quantitative metrics: frequency-domain entropy \( E = -\sum_i P_i \log(P_i) \), where \( P_i = \frac{A_i}{\sum_j A_j} \) and \( A_i \) represents the amplitude of the \(i^\text{th}\) FFT component, quantifies the richness of texture details,  higher \( E \) values indicate more diverse frequency content. The shift-invariance score is defined as \( S = \frac{1}{N}\sum_{k=-K}^{K} \text{SSIM}(I, I_k) \), where \( I_k \) is the FFT log-amplitude spectrum after horizontally shifting the input image \( I \) by \( k \) pixels, and \( N \) is the total number of shifts, SSIM refers to the structural similarity index \cite{wang2004image}. A higher \( S \) value implies greater robustness to spatial displacement.

In the original amplitude (3rd column), T1n and FLAIR show strong directional scattering and irregular fluctuations due to patient variation, scanner noise, and motion artifacts, which impair consistent global feature extraction. Taking the blurry T1n in the first row as a representative example, we observe that after DTCWT-based LF decomposition (4th column), the amplitude spectrum becomes notably smoother, with elevated LF baseline amplitudes reflecting enhanced global structural stability. This transformation markedly increased the shift-invariance metric $S$ (from 0.34 to 0.84), validating DTCWT’s inherent robustness. However, entropy $E$ slightly decreased (from 8.63 to 6.58), suggesting minor loss of HF texture. After integrating NSCT-based HF information (5th column), the fused amplitude became more uniformly distributed and higher overall, mitigating prior irregularities. Consequently, entropy $E$ rose significantly (to 10.19), indicating richer texture and structure, while $S$ remained substantially improved (0.71 vs. 0.34). Similar trends were consistently observed in other cases.

In summary, the proposed combination of DTCWT and NSCT  in the low and high frequency fusion strategy achieves a balance between global structural stability and detailed texture preservation, leading to more robust frequency representations and improved segmentation performance. However, the effectiveness of frequency decomposition relies on the assumption that directional and textural distinctions between tumor subregions are preserved across modalities. In real-world scans with severe inter-sequence misalignment or motion-corrupted high-frequency components, this assumption may be violated, potentially leading to degraded frequency separation and feature fusion.

\subsubsection{Ablation Analysis}
\mbox{}\\[2pt]
\hspace*{2em}\textbf{Ablation Study for Each Component in our HFF-Net.}
To assess the contribution of each module in a clinically realistic setting, we conducted ablation experiments on the BraTS2023-MEN dataset. As summarized in Table~\ref{tab:tab5}, introducing the FDD module yields strong baseline performance with Dice scores of 92.89\% (ET), 91.38\% (WT), and 92.48\% (TC), highlighting the effectiveness of frequency-based dual-branch encoding. Adding the ALC module further improves ET segmentation by 1.96\%, benefiting from its adaptive modeling of multi-directional HF details, while only marginally increasing parameters (by 0.72M) and FLOPs. Incorporating FDCA leads to notable gains, especially for TC (increases by 4.08\%), due to enhanced cross-slice semantic aggregation. The complete HFF-Net achieves the best overall performance, with Dice scores of 96.16\% for ET, 95.56\% for WT, and 96.34\% for TC, while maintaining a moderate complexity of 37.76 million parameters and 573.06 GFLOPs. These results confirm the effectiveness and complementarity of each component in enhancing segmentation accuracy.

\begin{table}[t]
    \centering
   
    \begin{subtable}
    \protect   \vspace{-18pt} \caption{Ablation Results for the Proposed Components on the BRATS2023-MEN Dataset }
    \label{tab:tab5}
    \resizebox*{1\linewidth}{0.076\textheight}{
    \begin{tabular}{c|ccc|cl|cl}
    \hline
     & \multicolumn{3}{c|}{\textbf{Dice Score (\%) ↑}} & \multicolumn{2}{l|}{{\color[HTML]{FFFFFF}}} & \multicolumn{2}{l}{} \\ \cline{2-4}
    \multirow{-2}{*}{\textbf{Method}} & \textbf{ET} & \textbf{WT} & \textbf{TC} & \multicolumn{2}{l|}{\multirow{-2}{*}{{\textbf{Param(M)}}}} & \multicolumn{2}{l}{\multirow{-2}{*}{\textbf{Flops(G)}}} \\ \hline
    FDD & 92.89 & 91.38 & 92.48 & \multicolumn{2}{c|}{37.01} & \multicolumn{2}{c}{564.97} \\
    FDD + ALC & 95.89 & 93.81 & 91.02 & \multicolumn{2}{c|}{37.73} & \multicolumn{2}{c}{569.93} \\
    {FDD + FDCA} & \multicolumn{1}{c}{94.85} & {93.73} & {95.56} & \multicolumn{2}{c|}{37.75} & \multicolumn{2}{c}{565.43}\\ \hline
    { \textbf{FDD + FDCA + ALC}} & {\color[HTML]{FE0000} \textbf{96.16}}& {\color[HTML]{FE0000} \textbf{95.56}}& {\color[HTML]{FE0000} \textbf{96.34}}& \multicolumn{2}{c|}{37.76} & \multicolumn{2}{c}{573.06} \\ \hline
    \end{tabular}}
    \end{subtable}

    \begin{subtable}
    \protect\caption{Ablation Study on Different Operators in ALC Convolutional Kernel Weights}
    \label{tab:tab6}
    \resizebox*{1\linewidth}{0.074\textheight}{
    \setlength{\arrayrulewidth}{0.1mm}
    \begin{tabular}{c|ccc|ccc}
    \hline
     & \multicolumn{3}{c|}{\textbf{Dice Score (\%) ↑}} & \multicolumn{3}{c}{\textbf{Hausdorff Dist. (mm) ↓}} \\ \cline{2-7} 
    \multirow{-2}{*}{\textbf{Operator}} & \textbf{ET} & \textbf{WT} & \textbf{TC} & \textbf{ET} & \textbf{WT} & \textbf{TC} \\ \hline
    Sobel & 91.35 & 91.72 & 89.50 & 6.372 & \color[HTML]{FE0000}\textbf{6.175} & 13.031 \\ 
    Scharr & 94.47 & 93.03 & 91.44 & 10.872 & 13.643 & 11.295 \\ 
    Kirsch & 95.58 & 92.81 & 93.29 & 7.283 & 9.381 & 7.58 \\ \hline
    \textbf{Discrete Laplace} & \color[HTML]{FE0000}\textbf{96.16} & \color[HTML]{FE0000}\textbf{95.56} & \color[HTML]{FE0000}\textbf{96.34} & \color[HTML]{FE0000}\textbf{5.576} & 6.272 & \color[HTML]{FE0000}\textbf{6.432} \\ \hline
    \end{tabular}}
    \ 
    \end{subtable}

    \begin{subtable}
    \protect\caption{Ablation Study on FDD Module: Decomposition Strategies, Levels, and Directions in HF Decomposition}
    \label{tab:tab7}
    \setlength{\arrayrulewidth}{0.2mm}
    \resizebox*{0.92\linewidth}{0.103\textheight}{
    
    \begin{tabular}{c|c|ccc|ccc}
    \hline
     &  & \multicolumn{3}{c|}{\textbf{Dice Score (\%) ↑}} & \multicolumn{3}{c}{\textbf{Hausdorff Dist. (mm) ↓}} \\ \cline{3-8}
    \multirow{-2}{*}{\textbf{Strategy(L/H)}} & \multirow{-2}{*}{\textbf{Levels,Directions}} & \textbf{ET} & \textbf{WT} & \textbf{TC} & \textbf{ET} & \textbf{WT} & \textbf{TC} \\ \hline
    DWT/DWT & [1,-] & 63.55 & 69.54 & 71.52 & 62.847 & 79.575 & 58.951 \\ \cline{2-2}
     & [1,6] & 88.34 & 87.07 & 85.32 & 18.205 & 21.934 & 23.397 \\
    \multirow{-2}{*}{DTCWT/DTCWT} & [2,6] & 81.50 & 82.03 & 83.59 & 29.587 & 34.043 & 29.359 \\ \cline{2-2} 
     & [1,4] & 92.82 & 93.38 & 91.36 & 10.548 & 9.832 & 10.285 \\ 
    \multirow{-2}{*}{NSCT/NSCT} & [2,4] & 87.54 & 88.31 & 85.46 & 24.659 & 19.512 & 21.539 \\ \cline{2-2}
     & [1,2] & 94.53 & 92.87 & 93.86 & 9.985 & 11.648 & 10.707 \\
     & \color[HTML]{FE0000}\textbf{[1,4]} & \color[HTML]{FE0000}\textbf{96.16} & \color[HTML]{FE0000}\textbf{95.56} & \color[HTML]{FE0000}\textbf{96.34} & \color[HTML]{FE0000}\textbf{5.576} & \color[HTML]{FE0000}\textbf{6.272} & \color[HTML]{FE0000}\textbf{6.432} \\
     & [2,2] & 91.31 & 90.55 & 89.78 & 11.921 & 16.322 & 19.805 \\
    \multirow{-4}{*}{\textbf{DTCWT/NSCT}} & [2,4] & 92.39 & 91.98 & 90.54 & 10.082 & 14.283 & 17.779 \\ \hline

    \end{tabular}}

\end{subtable}
 \vspace{-6pt} 
\end{table}

\begin{figure}[!t]
    \includegraphics[width=\columnwidth]{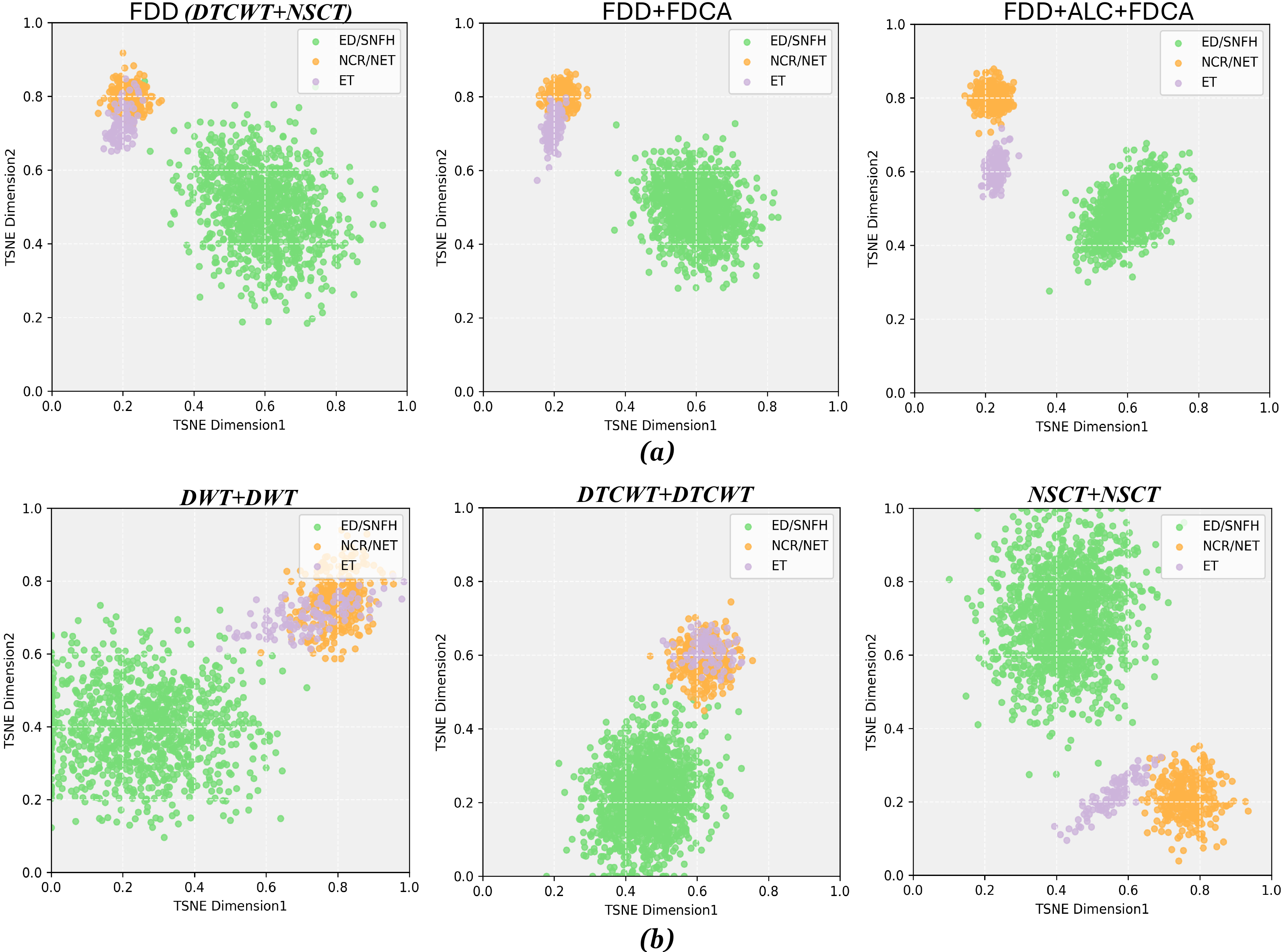} 
    \caption{ Feature visualization with t-SNE to show (a) how frequency domain decomposition (FDD), adaptive Laplacian convolution (ALC), and frequency domain cross-attention (FDCA) modules individually and jointly enhance feature representation; (b) comparison of alternative frequency decomposition combinations within the FDD module. Each color corresponds to a specific category of brain tumor sub-regions.}
    \label{Fig:tsne}
\vspace{-5pt}
\end{figure}

To further understand the contribution of each component to feature discrimination among tumor sub-regions, we conducted t-SNE \cite{van2008visualizing} visualizations on the extracted feature representations, as shown in \Cref{Fig:tsne} (a). When using only the FDD module (DTCWT+NSCT), feature distributions corresponding to peritumoral edema (ED/SNFH, green), necrotic or non-enhancing tumor (NCR/NET, orange), and enhancing tumor (ET, violet) show preliminary clustering. Nonetheless, intra-class cohesion is suboptimal, and ET and NCR/NET remain poorly separated. This suggests that frequency decomposition and fusion alone provide insufficient fine-grained semantic separability. With the addition of the FDCA module, each cluster becomes more compact, and inter-class boundaries are more clearly defined, indicating enhanced global structure encoding. However, some overlap between ET and NCR/NET still persists, reflecting limited discrimination in semantically complex regions. Finally, incorporating the ALC module yields the most distinct and cohesive clustering. In particular, ET and NCR/NET are effectively separated, demonstrating ALC’s ability to refine high-frequency, multi-directional features and improve recognition of intricate sub-regions such as enhancing tumor boundaries. However, in clinical scenarios where the contrast between enhancing and non-enhancing tumor regions is intrinsically low, such as in infiltrative gliomas or post-treatment scans which the frequency-based representation may still struggle to achieve sufficient separability. This indicates the potential need for integrating complementary semantic priors or uncertainty-aware modeling to enhance robustness.

 \textbf{Ablation Study Results for Different Discrete Operators Applied to ALC Convolutional Kernel Weights.} As presented in Table \cref{tab:tab6}, we evaluated the discrete Laplace operator along with three other HF edge detection operators \cite{gonzalez2017edge}. The discrete Laplace operator achieved the highest Dice scores (96.16\% for ET, 95.56\% for WT, and 96.34\% for TC) and the lowest HD95 distances (5.576 mm, 6.272 mm, and 6.432 mm, respectively). In comparison, the Sobel operator had Dice scores of 91.35\%, 91.72\%, and 89.50\%, with HD95 of 6.372 mm, 6.175 mm, and 13.031 mm. The Scharr operator improved Dice scores to 94.47\%, 93.03\%, and 91.44\%, but had higher HD95 (10.872 mm, 13.643 mm, 11.295 mm). The Kirsch operator performed moderately with Dice scores of 95.58\%, 92.81\%, and 93.29\%, and HD95 of 7.283 mm, 9.381 mm, and 7.58 mm.

\textbf{Ablation Study for Decomposition Strategy Combinations and HF Decomposition Layers and Directions in the FDD Module.} As shown in Table \cref{tab:tab7}, different combinations of LF and HF decomposition strategies significantly affect segmentation performance. The DWT/DWT setup yielded the lowest Dice scores and highest HD95, due to its inability to preserve spatial resolution and lack of directional decomposition. Using DTCWT for both LF and HF improved performance when configured with one decomposition level and six directions, but degraded when the level increased to two, highlighting DTCWT’s sensitivity to over-decomposition. The NSCT/NSCT configuration performed consistently well, with a [1,4] setting achieving a notable Dice improvement of up to 12.91\% (TC) and HD95 reduction of over 70\%, benefiting from NSCT's strong directional selectivity. Our proposed DTCWT/NSCT combination with [1,4] further improved performance across all regions, achieving Dice scores of 96.16\%, 95.56\%, and 96.34\% for ET, WT, and TC, respectively, and the lowest HD95 values. This confirms that using DTCWT for LF ensures spatial stability and shift-invariance, while NSCT for HF captures directional textures effectively. Moreover, configurations with one decomposition level consistently outperformed two-level setups, as excessive decomposition introduces noise in relatively low-resolution MRI volumes. Overall, balancing decomposition depth and orientation selectivity is essential for achieving accurate and robust segmentation.
To further validate the rationale and effectiveness of the proposed DTCWT-NSCT combination, we conducted a t-SNE analysis to visualize the feature distributions generated by different frequency decomposition combinations, as shown in \Cref{Fig:tsne} (b). Using DWT for both LF and HF components resulted in highly dispersed feature points with significant overlap between enhancing tumor (ET) and necrotic/non-enhancing tumor (NCR/NET), due to DWT’s lack of directional and scale selectivity. In contrast, DTCWT–based decomposition yielded more compact clusters, benefiting from shift-invariance, though ET and NCR/NET remained poorly separated. Applying NSCT to both bands led to improved distinction between ET and NCR/NET, thanks to its multi-directional decomposition capability that captures fine texture and edge details, albeit with slightly looser clustering for peritumoral edema (ED/SNFH). Notably, our DTCWT–NSCT hybrid strategy achieved the most compact and clearly separated clusters across all sub-regions, effectively balancing structural stability and fine-grained semantic discriminability. This supports the superiority and rationale of our frequency-domain design in the FDD module.

 \begin{figure}[!t]
    \includegraphics[width=\columnwidth]{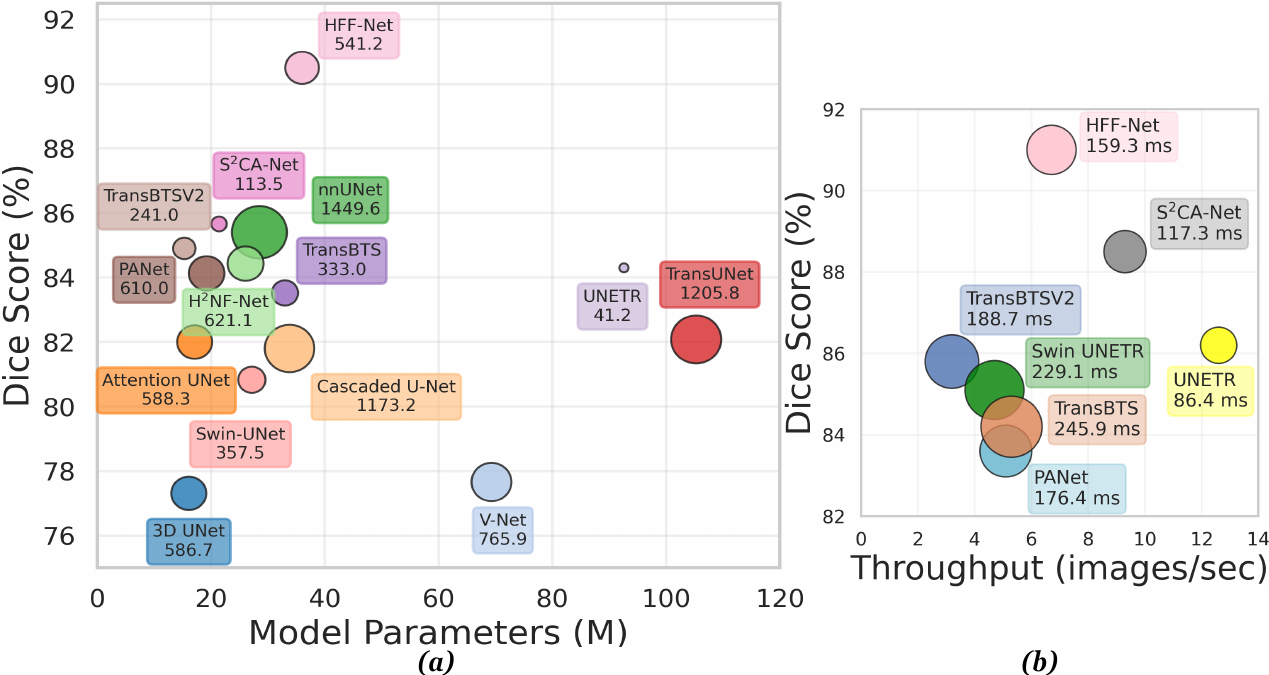} 
    \caption{Comparative visualization of trade-offs among methods regarding computational complexity, clinical applicability, and segmentation performance. (a) Depicts computational complexity using parameter size, Dice score, and GFLOPs (bubble size). (b) Illustrates inference efficiency with throughput (images per second), Dice score, and GPU processing time (bubble size).}
    \label{Fig:ce}

\end{figure}








\subsubsection{Computational Complexity and Efficiency Analysis}
\mbox{}\\[2pt]
\hspace*{2em}\textbf{Computational Complexity and Segmentation Performance.}
To assess the practical applicability of our method in clinical scenarios, we analyzed the trade-off between model complexity and segmentation performance. As shown in \Cref{Fig:ce} (a), HFF-Net achieves the highest Dice score (90.50\%) among all compared methods, despite having a moderate number of parameters (36.01M) and computational cost (541.2 GFLOPs). Notably, it outperforms models with significantly larger parameter sizes, such as TransUNet (105.28M) and nnUNet (28.5M), which achieve only 82.08\% and 85.40\% Dice, respectively. In contrast, lightweight models like S$^2$CA-Net and TransBTSV2 exhibit lower complexity but still lag behind in accuracy, highlighting the effectiveness of HFF-Net’s dual-branch frequency-aware architecture and attention mechanisms in utilizing model capacity efficiently. Each bubble in the plot represents a method, with its color uniquely assigned and its size proportional to GFLOPs, offering a visual depiction of computational demand. While HFF-Net balances performance and complexity effectively, its relatively higher resource requirements may limit deployment in resource-constrained clinical settings. Future work may focus on optimizing the architecture to reduce complexity while preserving segmentation quality.

\textbf{Inference Efficiency and Clinical Applicability.}
In time-sensitive clinical environments, inference efficiency plays a crucial role in supporting diagnostic decisions and real-time interventions. As illustrated in \Cref{Fig:ce} (b), HFF-Net achieves a throughput of 6.71 images per second and a GPU processing time of 159.3 ms per image, performing favorably compared to models such as Swin UNETR (229.1 ms) and TransBTS (245.9 ms). Although it is marginally less efficient than lightweight architectures like S$^2$CA-Net (117.3 ms) and UNETR (86.4 ms), the substantial gain in accuracy justifies the computational trade-off in many clinical contexts. The plot encodes GPU latency as bubble size and assigns a distinct color to each model for clarity. Overall, HFF-Net demonstrates a strong balance between precision and efficiency, which can reduce the need for manual correction and accelerate clinical workflows. Nevertheless, for applications with strict real-time constraints, such as intraoperative navigation or emergency diagnosis, further improvements in inference speed are desirable to enhance its clinical versatility.

 \begin{figure}[!t]
    \includegraphics[width=\columnwidth]{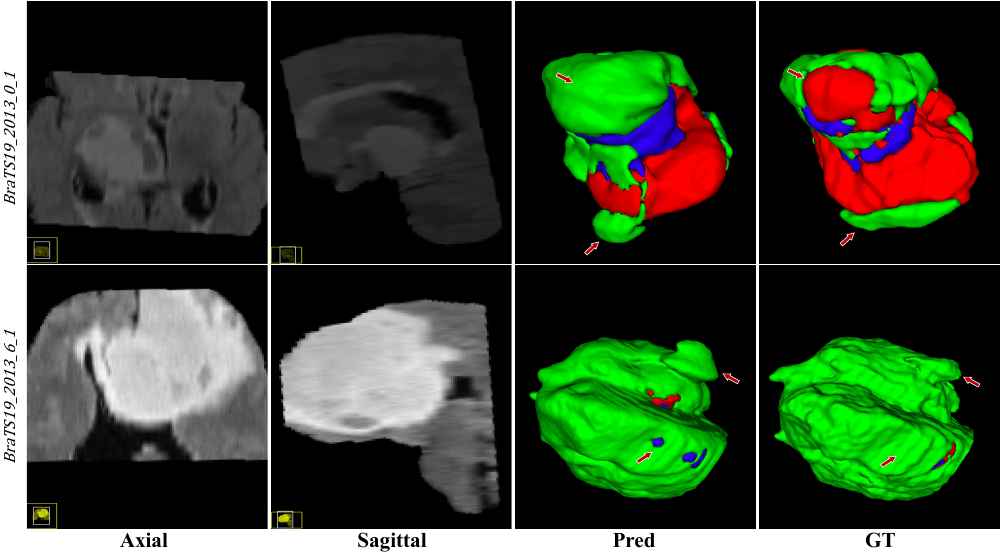} 
    \caption{Visualization of failure cases to show how missing FLAIR modality signals affect segmentation. \textcolor[HTML]{FE0000}{Red} arrows highlight areas of significant deviation from the ground truth.}
    \label{Fig:fc}

\end{figure}
\subsubsection{Failure Case Analysis}
While HFF-Net achieves strong overall performance, failure cases still occur under conditions of abnormal or incomplete modality signals. As illustrated in \Cref{Fig:fc}, we present two examples where segmentation errors arise due to partial corruption of the FLAIR modality. In both cases, necrotic or enhancing tumor regions (red and blue) were misclassified as peritumoral edema (green). Such misclassifications occur because the FLAIR modality plays a pivotal role in highlighting edema with high signal intensity, enabling clear discrimination from necrotic or non-enhancing regions. In contrast, T1 and T2 modalities alone exhibit limited contrast for this purpose, making it difficult for the model to reliably resolve the boundary between these subregions. This modality dependence becomes particularly problematic when FLAIR signals are partially missing, blurred, or corrupted, as observed in cases affected by motion artifacts, field-of-view truncation, or acquisition anomalies.

From a representational standpoint, frequency decomposition in the FDD module is modality-agnostic and applied uniformly to all inputs. Consequently, when a key modality like FLAIR is degraded, the decomposed frequency components may no longer accurately reflect the expected anatomical structure, resulting in ambiguity during both training and inference. Moreover, as shown in the visualizations, the network fails to preserve semantic separation between structurally similar but pathologically distinct regions, particularly in volumetric renderings, where misclassified edema occupies spatially inconsistent and morphologically implausible zones.

These failure cases underscore the need for future models to incorporate modality-aware mechanisms that can explicitly detect and compensate for corrupted or missing input signals. Potential strategies include uncertainty modeling, attention masking, or plug-in robustness modules that learn adaptive weighting schemes across modalities based on their reliability. Integrating such mechanisms may enhance the generalizability of frequency-aware networks in realistic clinical settings, where imperfect imaging data is the norm rather than the exception.
\begin{table}[t]
\centering
\caption{Broad-impact quantitative analysis comparing classic UNet and a 2D HFF-Net variant across diverse medical imaging modalities.}
\label{tab:2d_mixed_metrics}
\resizebox{0.8\linewidth}{!}{%
  \begin{tabular}{c|cc|cc}
    \hline
    \multirow{2}{*}{\textbf{Dataset}} & \multicolumn{2}{c|}{\textbf{UNet}} & \multicolumn{2}{c}{\textbf{Ours (2D)}} \\
    \cline{2-5}
    & \textbf{Dice (\%)↑} & \textbf{IoU (\%)↑} & \textbf{Dice (\%)↑} & \textbf{IoU (\%)↑} \\
    \hline
    CREMI        & 85.57 & 75.22 & 89.37 & 80.59 \\
    GlaS         & 88.24 & 81.58 & 92.83 & 85.41 \\
    ISIC2018     & 85.26 & 77.53 & 91.90 & 84.27 \\
    Kvasir-SEG   & 81.20 & 74.63 & 92.16 & 89.43 \\
    \hline
  \end{tabular}%
}
\vspace{-3pt}
\end{table}

 \begin{figure}[!t]
 \centering
    \includegraphics[width=0.75\columnwidth]{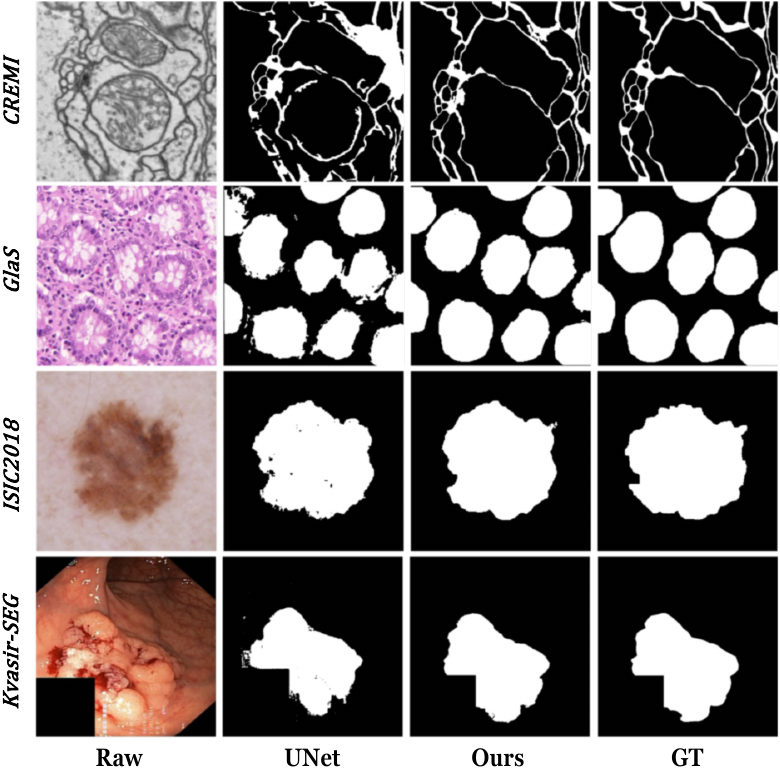} 
    \caption{Cross-domain segmentation on four 2D datasets shows that our frequency-decoupled design generalizes across medical imaging modalities by maintaining structural integrity and enhancing fine-grained boundaries.}
    \label{Fig:ba}

\end{figure}

\subsubsection{Broader Impact Analysis}
To further assess the generalizability and broader applicability of our frequency-decoupled framework, we implemented a two-dimensional variant of HFF-Net and evaluated it on four public 2D medical image segmentation datasets, each representing a distinct imaging modality. Specifically, we selected the CREMI dataset \cite{CREMI2016} for neuronal membrane segmentation from electron microscopy images, the GlaS dataset \cite{sirinukunwattana2017gland} for gland segmentation in hematoxylin and eosin stained histopathological slides, the ISIC2018 dataset \cite{codella2019skin} for skin lesion segmentation from dermoscopic photographs, and the Kvasir-SEG \cite{jha2019kvasir} dataset for polyp segmentation in colonoscopy images. For the CREMI and GlaS datasets, we adopted the official training and testing protocols as defined in prior benchmarks. For ISIC2018 and Kvasir-SEG, we randomly divided the data into 80\% training and 20\% testing subsets. During training and inference, all input images were resized to $256 \times 256$, and standard augmentations (flipping, rotation, transposition) were applied. Performance on all datasets was reported using Dice and IoU \cite{jaccard1901etude} metrics for consistency and comparability.

To ensure fair optimization across datasets, we used dataset-specific hyperparameter settings. For GlaS, the initial learning rate was set to 0.4, with $\lambda_{\text{max}} = 6$ and a batch size of 2. For CREMI, we used a learning rate of 0.5, $\lambda_{\text{max}} = 1$, and a batch size of 2. For ISIC2018 and Kvasir-SEG, the initial learning rate was set to 0.5, $\lambda_{\text{max}} = 2$, and the batch size was fixed at 2. All models were optimized using stochastic gradient descent (SGD) with momentum 0.9 and a weight decay of $5 \times 10^{-5}$. The 2D variant of HFF-Net retains the core frequency-aware architecture, replacing all 3D convolutions with their 2D counterparts and reducing the input to a single modality. Additionally, the FDCA module was adapted by omitting inter-slice attention components to suit the 2D setting. 

As shown in Table~\Cref{tab:2d_mixed_metrics} and ~\Cref{Fig:ba}, the 2D adaptation of our method consistently surpasses baseline UNet across four heterogeneous datasets, spanning electron microscopy, histology, dermoscopy, and endoscopy. On CREMI and GlaS, our approach yields more coherent boundary localization and lower HD95, reflecting improved modeling of anisotropic and topology-sensitive structures. On ISIC2018 and Kvasir-SEG, it achieves superior lesion separation and contour regularity under diverse appearance conditions. These findings reinforce the medical imaging modality agnostic and structurally adaptive nature of our frequency-aware architecture, validating that its benefits are not confined to volumetric MRI or neuro-oncology applications. Instead, the explicit decomposition and disentanglement of directional and spectral representations serve as a general strategy to address common challenges in medical segmentation, namely low-contrast ambiguity, irregular morphology, and boundary discontinuity. More broadly, this suggests that frequency-domain modeling may offer a foundational design principle for next-generation segmentation frameworks, capable of generalizing across anatomical targets, imaging protocols, and clinical use cases.

\section{Conclusion and Future Work} 
\label{sec:sec5}

In this work, we proposed HFF-Net, a segmentation framework that leverages frequency-domain decomposition to effectively capture the complex textures and directional structures inherent in brain tumor MRI. Although our model demonstrates significant improvement in segmenting contrast-enhancing tumor regions by explicitly integrating low- and high-frequency features, challenges remain regarding robustness to partially corrupted or incomplete imaging modalities, as highlighted by our failure analysis. Future research could explore advanced approaches such as modality-adaptive or missing-modality-aware strategies to address these limitations, potentially incorporating uncertainty modeling to enhance reliability in clinically realistic scenarios. Additionally, while HFF-Net achieves a strong balance between segmentation accuracy and computational efficiency, optimizing the model architecture to further reduce computational demands would significantly facilitate its deployment in resource-limited clinical environments. Extending this frequency-domain approach beyond brain MRI to other modalities and anatomical structures presents another promising avenue, potentially broadening its clinical utility and impact across diverse medical imaging applications.

This study situates frequency-domain modeling within the broader landscape of medical image segmentation, offering a principled alternative to prevailing spatial-domain approaches. By integrating classical signal decomposition techniques into a deep learning framework, our method complements existing architectures with enhanced representation of anatomical texture and directionality. Beyond performance gains, this work contributes a novel perspective that expands the design space of segmentation methods, highlighting the continued relevance of frequency analysis in achieving robust, interpretable, and modality-adaptive solutions.

\section{Acknowledgements}

This work was supported by the National Natural Science Foundation of China [grant number 62401097]; and the Natural Science Foundation of Liaoning Province (Doctoral Research Startup Project) [grant number 2024-BS-028]; and the Scientific Research Funding Program of Liaoning Provincial Department of Education [grant number LJKZ0024]; and the Fundamental Research Funds for the Central Universities, Dalian Minzu University [grant number 0854-53].

\bibliographystyle{IEEEtran}
\bibliography{ref}

\end{document}